\documentclass{aa}  

\usepackage{graphicx}
\usepackage{enumerate}
\usepackage[breaklinks=true]{hyperref} 
\bibpunct{(}{)}{;}{a}{}{,} 
\bibpunct{(}{)}{;}{a}{}{,}
\usepackage{graphicx}
\usepackage{txfonts}
\usepackage{lscape}
\usepackage{siunitx}
\usepackage{hyperref}
\begin{document} 
 \title{Westerlund 1 under the light of \textit{Gaia} EDR3: \linebreak
 Distance, isolation, extent, and a hidden population\thanks{"Tables B1, C1 and C2 are only available in electronic form at the CDS via anonymous ftp to \url{cdsarc.u-strasbg.fr} (\url{130.79.128.5})
or via \url{http://cdsweb.u-strasbg.fr/cgi-bin/qcat?J/A+A/}}.
}  
   \author{I. Negueruela \inst{1}
          \and E.~J. Alfaro \inst{2}
          \and R. Dorda \inst{3,4}
          \and A. Marco \inst{5}
          \and J. Ma\'{\i}z Apell\'aniz \inst{6}
          \and C. Gonz\'alez-Fern\'andez \inst{7}}
  \institute{Departamento de F\'{\i}sica Aplicada, Facultad de Ciencias, Universidad de Alicante, Carretera de San Vicente s/n, E03690, San Vicente del Raspeig, Spain
\email{ignacio.negueruela@ua.es}
\and
Instituto de Astrof{\'i}sica de Andaluc\'{\i}a, CSIC, Glorieta de la Astronomía s/n, E18008, Granada, Spain  
\and
Instituto de Astrof\'{\i}sica de Canarias,  V\'{\i}a L\'actea s/n, E38200, La Laguna, Tenerife, Spain
\and
 Universidad de La Laguna, Departamento de Astrofísica, E38206 La Laguna, Tenerife, Spain
 \and
 Departamento de F\'{\i}sica, Ingenier\'{\i}a de Sistemas y
Teor\'{\i}a de la Se\~{n}al, Universidad de Alicante, Carretera de San Vicente s/n, E03690, San Vicente del Raspeig, Spain
\and
Centro de Astrobiolog\'{\i}a, CSIC-INTA. Campus ESAC. 
              C. bajo del castillo s/n. 
              E-\num{28692} Villanueva de la Ca\~nada, Madrid, Spain
\and
Institute of Astronomy, University of Cambridge, Madingley Road, Cambridge CB3 0HA, UK}
   \offprints{}

   \date{Received ; accepted}
\titlerunning{Size of Wd~1}

   \date{}

 
  \abstract
 {The young massive cluster Westerlund~1 offers the promise of a grand laboratory for the study of high-mass star evolution, but its basic parameters are still poorly known.}
   {In this paper, we aim at a better characterisation of the cluster by determining some basic kinematic properties and analysing the area surrounding the cluster and the population in its foreground.}
   {We have used \textit{Gaia} early data release 3 (EDR3) data, together with spectra of a large sample of luminous stars in the field surrounding Westerlund~1, to explore the extent of the cluster. We carried out a non-parametric analysis of proper motions and membership determination. We investigated the reddening and proper motions of several dozen OB stars and red supergiants less than one degree away from Westerlund~1.}
   {We identify a population of kinematic members of Westerlund~1 that largely includes the known spectroscopic members. From their EDR3 parallaxes, we derive a distance to the cluster of $4.23^{+0.23}_{-0.21}$~kpc. We analyse the extinction in this direction, finding that it increases by a large amount around 2.8~kpc, which in all likelihood is due to dark clouds associated with the Scutum-Crux arm. As a consequence, we hardly see any stars at distances comparable (or higher) than that of the cluster. The proper motions of Westerlund~1, however,  are very similar to those of stars in the field surrounding it which are -- almost without exception -- less distant, but distinct. We find a second, astrometrically well-defined population in the foreground ($d\approx2\:$kpc), centred $\sim8\arcmin$ away, which is likely connected to the possible open cluster BH~197. Westerlund~1 is very elongated, an effect that seems real and not driven by the very heavy extinction to the east and south. We find a low-density halo extending to distances up to $10\arcmin$ from the cluster centre, mainly in the north-west quadrant. A few OB stars at larger distances from the cluster, most notably the luminous blue variable (LBV) MN48, share its proper motions, suggesting that Westerlund~1 has little or no peculiar motion with respect to the field population of the Norma arm. Despite this, we are unable to find any red supergiant that could belong to an extended population related to the cluster, although we observe several dozen such objects in the foreground, demonstrating the richness of the field population along this sightline. We find a substantial population of luminous OB members obscured by several more magnitudes of extinction than most known members. These objects, mostly located in the central region of the cluster, increase the population of OB supergiants by about 25\%.}
   {}


   \maketitle
%

\section{Introduction}
Westerlund~1 (Wd~1) is believed to be the most massive young cluster in the Milky Way, with mass estimates ranging from about $5\times10^{4}$ to $\ga10^{5}\:\mathrm{M}_{\sun}$ \citep{clark05,Brandner2008,gennaro11}. Beyond the anecdotal interest of being a top contender for the title of most massive young cluster in the Local Group -- in all likelihood beaten by R136 in the Large Magellanic Cloud -- the huge mass of Wd~1 turns it into a prime laboratory for massive star evolution. There are so many massive stellar systems in the cluster that rare evolutionary phases are sampled, and different avenues for binary interaction are probed, resulting in a variety of outcomes \citep[see, e.g.][]{Clark11WC}.

The population associated with the cluster encompasses all sorts of massive stars, from O-type giants and supergiants to red supergiants \citep{clark20}, including a number of extremely luminous B-type hypergiants \citep{negueruela10Wd1}, an LBV \citep{ritchie09W243}, several yellow hypergiants \citep{clark10}, some transitional B-type supergiants with emission lines \citep{ritchie10, Clark14W5}, an extremely luminous B[e]-like object \citep{clark13W9}, and a large population of Wolf-Rayet stars \citep{crowther06WRs}. The vast majority of these objects lie within less than 1 arcmin from the nominal cluster centre, although some outliers, such as the Wolf-Rayet stars known as N (= WR 77n) or X (= WR 77sd), are more than 3 arcmin away. 

All these varieties of massive objects are generally assumed to correspond to stars with initial masses $\ga30\,M_{\sun}$ and imply ages $\la 5$~Ma for stars evolving in isolation (if binary evolution is considered, the age is somewhat older). Nevertheless, an empirical determination of their masses will depend, to a large degree, on the parameters derived for the cluster. Observed stellar magnitudes are transformed into a luminosity and hence a mass, impacting heavily on the age adopted for Wd~1. The two main unknowns are the true distance to Wd~1 and the effect of extinction on the determination of luminosities. The cluster is affected by heavy reddening. Different estimates of the average extinction towards Wd~1 encompass the $A_{V}\approx10$\,--\,12 range, although there are substantial variations across the face of the cluster \citep{negueruela10Wd1,damineli16}.

With such high extinction, derivation of absolute magnitudes is uncertain, not only because of the possibility of a non-standard extinction law, but also because of the extreme colour terms involved in the colour transformations \citep{clark05}, which may lead to systematic errors. Nevertheless, \citet{negueruela10Wd1} found no strong evidence for large deviations from a standard extinction law from the analysis of $VRI$ photometry. Contrarily, the analysis of near-IR data suggests a much steeper extinction law \citep{Lim2013, damineli16}.

Under the assumption of a standard extinction law, different estimates based on the observed massive star population favour distances around 5~kpc for Wd~1 \citep{clark05,crowther06WRs,negueruela10Wd1}. Conversely, the analysis of the near-IR colour magnitude diagram with pre-main sequence isochrones by \citet{Brandner2008} resulted in a lower distance of 3.6~kpc. A close value of 3.8~kpc was also obtained from NIR photometry by \citet{Lim2013}. \citet{kothes07} obtained an indirect estimate of the cluster distance by resorting to the gas clouds in its immediate neighbourhood. By assuming that the clouds were physically connected to Wd~1 and that they followed the Galactic rotation curve, they determined $d\approx3.9\pm0.7$~kpc. Although these are sensible assumptions, the  Galactic spiral  pattern delineated by maser radial velocities and trigonometric parallaxes  \citep{Reid2014, Reid2019} shows a very complex situation along this sightline ($l \approx 340\degr$), where the  Scutum--Centaurus and Norma--Outer arms are not easily distinguishable, and tracers of a given arm show a large dispersion in velocity.

\textit{Gaia}~DR2 data for the region of Wd~1 were affected by very strong systematic effects, which rendered any distance estimation very unreliable \citep{clark20}. \citet{aghakhan20} claimed to be able to measure an accurate parallax to the cluster from a Bayesian analysis of parallaxes to stars along the line of sight, obtaining a value of $\pi =0.35^{+0.07}_{-0.06}$~mas. Such a result is surprising, given that the median parallax for known members is 0.19~mas \citep{clark20}, and implies a distance of only $2.6^{+0.6}_{-0.4}\:$ kpc. This value is substantially shorter than any previous determination and would, according to these authors, imply a total mass not much higher than $2\times10^{4}\:\mathrm{M_{\sun}}$. \citet{davbeas19} carried out a more careful analysis, by selecting only stars whose proper motions were similar to those of known cluster members, coming to a distance of ${3.9}^{+1.0}_{-0.64}\:$kpc, more in line with previous authors. 

More recently, \citet{aghakan21} have insisted on a distance of ${2.8}_{-0.6}^{+0.7}\:$kpc derived from an analysis of the distribution of parallaxes in \textit{Gaia}~EDR3 for all the stars in the area, without any reference to membership. Meanwhile, \citet{beasor21} have used EDR3 data on OB stars with a coherent proper motion distribution to derive a substantially longer distance of 4.1$_{-0.36}^{+0.66}\:$kpc. Moreover, based on an estimation of the luminosity of the cool supergiants in Wd~1, \citet{beasor21} have argued that all these objects have ages around 10~Ma, implying masses of only $\sim17\:\mathrm{M}_{\sun}$, while the presence of some more massive stars and the younger age derived from pre-main-sequence isochrone fits can be explained by non-coeval or extended star formation.

The aim of this paper is to provide a well-founded estimation of the extent of Westerlund~1 and its possible connection to other populations in its surroundings, with the purpose of evaluating the likelihood of a complex (multi-age or multi-cluster) population. For this, we use two complementary tools: \textit{Gaia}~EDR3 astrometric data and spectroscopy of a large sample of luminous stars in a field of radius one degree surrounding the cluster. After introducing the data used in Section~\ref{sec:data}, we carry out membership analysis on the \textit{Gaia} EDR3 data in Sect.~\ref{sec:members} and then proceed to calculate an accurate distance to the cluster in Sect.~\ref{sec:distance}. We then describe the population surrounding the cluster in Sect.~\ref{sec:spectra} and explore the consequences of our findings in the Discussion. We close the paper with our conclusions.

\section{Data collection}
\label{sec:data}

\subsection{Spectroscopy data}
\label{subsec:spectroscopicdata}
Observations of the field surrounding Westerlund~1 were obtained with the fibre-fed dual-beam AA$\Omega$ spectrograph  mounted on the 3.9\,m Anglo  Australian Telescope (Siding Springs, Australia) on the nights of 2011, July 20 and 21. The Two Degree Field (2dF) multi-object system was utilised to position fibres. The instrument allows simultaneous observations in two different arms by using a dichroic beam-splitter with crossover at 5\,700\,\AA. Each arm of the AAOmega system was equipped with a 2k$\times$4k E2V CCD detector and an AAO2 CCD controller. In the blue arm, we used grating 580V, which gives a resolving power $R=1\,300$ over $\sim2100$\,\AA\ (with central wavelength at 4\,500\AA). Given the extremely high extinction to most of our targets, useful blue spectra were obtained for only a handful of objects. In the red arm, the 1700D grating was used, providing $R\approx11\,000$ in a 500\,\AA\ wide spectral window containing the near-infrared \ion{Ca}{ii} triplet (CaT). This grating must be centred on 8700\,\AA, but the actual central position of each spectrum is determined by the position of the target in the sky, with a maximum shift of 20\,\AA.

The targets observed were selected by means of a combination of photometric catalogues, following the procedures outlined in \citet{ns07} and \citet{negueruela12}, by means of the  $Q_{\mathrm{IR}}$ index, defined as $Q_{\mathrm{IR}}$\, =\, $(J-H) - 1.8 \cdot (H-K_{\mathrm{S}})$. A sample of "early-type" stars was created by selecting objects from the 2MASS \citep{skru06} catalogue with \textit{good} flags (i.e.\ "A" or "E") in all filters whose $Q_{\mathrm{IR}}$ index falls in the range between $-0.5$ and $+0.1$, which is occupied by emission-line stars and early (OBA) stars. Some F stars may also be included, as room is made for photometric errors, but the main contaminant population are late-M AGB stars, whose SEDs are similar to those of reddened OB stars \citep[e.g.][]{maiz20}. This initial selection was then cross-matched with the DENIS $i$ band (or USNO-B1.0 $I$ band, if DENIS data did not exist) to guarantee that the target would be observable ($i<15$). The list was completed with the very few catalogued early-type stars in the field and known members of Wd~1 (which, in most cases, have no reliable 2MASS magnitudes due to crowding and saturation).

Candidate cool luminous stars were selected with the same criteria that had worked successfully in other highly extincted areas \citep{negueruela12}. We picked very bright stars in the infrared ($K_{\mathrm{S}}<7$), with $E(J-K_{\mathrm{S}})>1.3$ (i.e. redder than the intrinsic colour of any red supergiant) and $Q_{\mathrm{IR}}< 0.4$ (a cut that leaves out most red giants). Again, a cross-match with DENIS and USNO-B1.0 was conducted to make sure that the star would be observable. Known cool supergiant members of Wd~1 were added by hand. 

On the night of July 20, we observed a \textit{bright} field, aimed at stars with $I<11.5$, with three exposures of 300\,s. The following night, we took four exposures of 1200\,s, targeting mostly fainter targets. We used the standard reduction pipeline {\tt 2dfdr} as provided by the AAT at the time. Wavelength calibration was attained by observing arc lamps before each target exposure. Details of the reduction can be found in \cite{gonzalez2015}.

\subsection{\textit{Gaia} astrometric and photometric data}
\label{subsec:Gaiadata}

\textit{Gaia}~EDR3 has provided unique quality photometric and astrometric data for more than 1.5 billion celestial objects \citep{GaiaEDR3}. The precision of the astrometric data (parallax and proper motions) is over 20\% better, on average, than in the previous release, \textit{Gaia}~DR2. We have made use of the version of the EDR3 catalogue available from Vizier (I/350/gaiaedr3) to obtain astrometry and photometry for the stellar sample analysed in this work. We downloaded data for all stars within a circle of radius 12 arcmin around the nominal centre of Wd~1 given by the SIMBAD database (ICRS; Epoch 2000; RA = $251.76667\degr$;\ DEC = $-45.85136\degr$; \citealt{Brandner2008}). The large radius chosen, more than six times larger than the 1.7 arcminutes catalogued by \cite{sampedro17}, is recommended by our objective of analysing the cluster and its environment in a highly crowded region of the sky whose line of sight could be crossing several spiral arms, and, therefore, a variety of stellar populations with different kinematic properties  \citep[e.g.][]{Reid2014, Reid2019}.

The selection has only been filtered by the condition that the re-normalized unit weight error (RUWE) is $<$\ 1.4.  In addition, data are only taken from a proper motion box defined by  $\mu_{\alpha}$ [-10; 7] and $\mu_\delta$ [-11; 5] in mas/a units\footnote{In what follows, $\mu_{\alpha}$ corresponds to $\mu_{\alpha}\cos(\delta)$ in mas/a.}. RUWE is a quality-fit parameter  of  Gaia  astrometric solutions \citep{lindegren21_astro}. It is directly linked to the $\chi^2$ statistic. Several authors suggest a RUWE threshold at 1.4 for the selection of large data samples, as this upper limit guarantees a well-behaved astrometric solution. This would be equivalent to choosing  objects with $\chi^2\,\leq\, 2$ \citep[e.g.][among others]{Stassun2021}.  Here, we adopt this criterion. The proper motion box is defined to remove the most obvious outliers in the proper motion space to prevent them from biasing the selection algorithm. The number of stars in the sample selected is 19\,432. Several authors \citep[e.g.][]{Lindegren2021, maiz22, Riello2021, Huang2021, Yang2021, Niu2021, CG2021}, have discussed the possible systematic errors that could still be present in astrometry and photometry published in \textit{Gaia} EDR3. We return to these corrections throughout the work, when we consider that their application could affect the  physical variables estimated, and/or modify some selection criteria or some of the conclusions reached.

\section{Membership analysis}
\label{sec:members}
 
\subsection{Gaia~EDR3: Membership exploratory analysis}

\begin{figure}
   \resizebox{\columnwidth}{!}
   {\includegraphics[angle=0, clip]{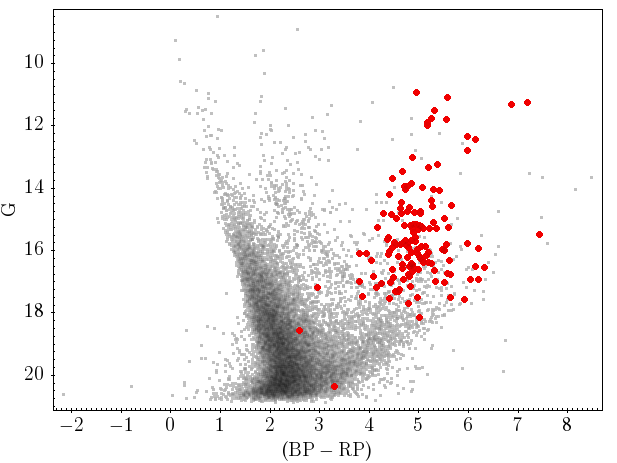}}
   \caption{$G$\  vs\  $(BP-RP)$ colour-magnitude diagram for the total sample with RUWE $\leq$\ 1.4 (grey small dots). Stars with spectroscopic information in \citet{clark20} and \textit{Gaia}~EDR3 photometry are shown in red. This sample is clearly seen to consist mostly of stars redder than $(BP-RP) >$\ 4.}
\label{fig:CMD3B}
\end{figure}

As mentioned in the Introduction, the stellar population associated with Wd~1 contains highly reddened, intrinsically very luminous, massive stars. With this prior knowledge, we explore how \textit{Gaia}~EDR3 photometry informs us about the stellar population in the Wd~1 field. Figure~\ref{fig:CMD3B} shows the $G$\ vs\  $(BP-RP)$ diagram for the full sample. We have marked in red the 134 stars for which spectral types have been provided by \citet{clark20} and \textit{Gaia}~EDR3 photometry exists in our sample. From all previous analyses \citep[e.g.][and references therein]{clark20}, all these objects are very likely cluster members. More than 90\% of these stars have a $(BP-RP)$ colour redder than 4~mag, with the remainder having very slightly bluer values. For an initial exploration of cluster parameters, we chose stars with $(BP-RP)\ >$\ 4, according to their \textit{Gaia}~EDR3 photometry. We call these objects the Red Branch (RB) stars. It is noteworthy that any correction to the \textit{Gaia}~EDR3 photometry, as suggested by some authors \citep[][and references therein]{Yang2021, Niu2021}, is too small to alter in any significant way the  sample of RB stars selected.


\begin{figure}
   \resizebox{\columnwidth}{!}
   {\includegraphics[angle=0, clip]{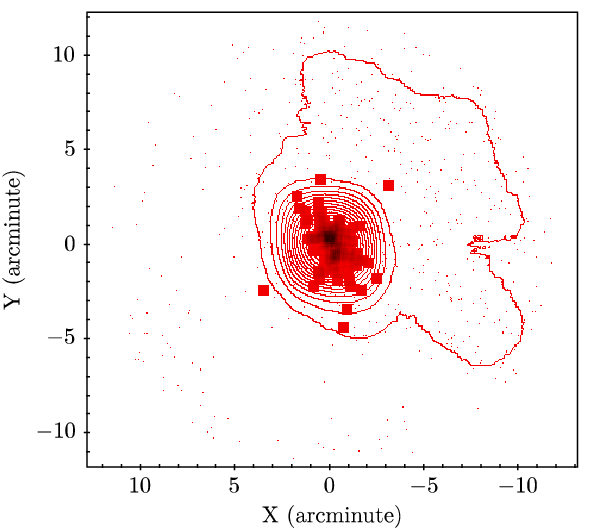}}
   \caption{Red Branch stars, defined as those with $(BP-RP)\ >$\ 4, plotted on the plane of the sky (dots). Coordinates are angular distance to the cluster centre along RA (X axis) and DEC (Y axis). Red squares identify stars with spectra reported in the catalogue of \citet{clark20}. The vast majority of red squares are within a radius of $\approx3\farcm5$.}
   \label{fig:XY_RB}
    \end{figure}

As a first approach, we explored how this data set behaves in different subspaces of the phase space formed by the astrometric variables ($\alpha,  \delta, \mu_{\alpha}, \mu_{\delta}, \varpi$). Initially, we did not use any parallax correction, as we were simply trying to determine the shape and structure of the distributions. For a more quantitative analysis, we later apply the necessary systematic corrections to parallax at the cost of substantially reducing the sample.
In Fig.~\ref{fig:XY_RB}, we show the spatial distribution of the 975 RB stars in the field of Wd~1. The coordinates are angular distances to the centre of the cluster in RA (X) and DEC (Y), both  measured in arcminutes. The stars distribute forming an elliptical and very dense central core whose major axis is $\approx$\ $7\arcmin$ long. Surrounding the core, a halo can be seen, mainly in the north-west quadrant, with a density of objects much lower (a factor 50) than the core density maximum. 

\begin{figure}
   \resizebox{\columnwidth}{!}
   {\includegraphics[angle=0, clip]{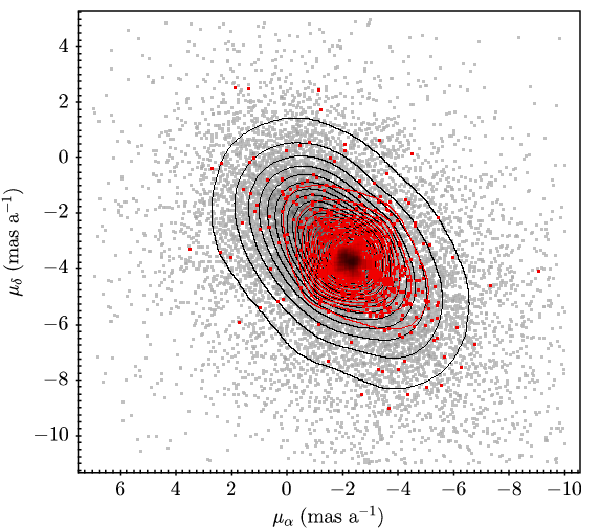}}
   \caption{Isodensity contours (black line) of the total sample (grey small dots). Red dots are RB stars, 
   which are in an eccentric position with respect to the whole sample distribution. They appear centred on one focus of the internal dispersion ellipse.}
   \label{fig:VPD_RB}
    \end{figure}

The \textit{Vector Point Diagram} (VPD) in Fig.~\ref{fig:VPD_RB} displays a similar pattern. Here, we also include the complete EDR3 sample to show how the distribution of the RB stars differs from that of the
bulk of data. The distribution of the whole sample is more elongated and centred on a nearby, but significantly different, position to the centre of the RB stars.

\begin{figure}
   \resizebox{\columnwidth}{!}
   {\includegraphics[angle=0, clip]{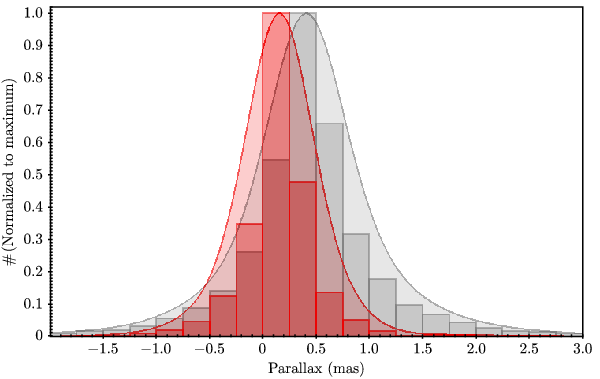}}
   \caption{Histogram showing the parallax distribution for the total sample (grey) and the RB stars (red). Frequency is normalised to the maximum value for better visualising the difference in modes. Attending to the histogram,  this difference is  0.25 mas (bin size). Kernel density estimators lead to a slightly shorter separation of $\approx$\ 0.20 mas.}
   \label{fig:Plx_RB}
    \end{figure}

The histogram of the parallaxes for the total sample and for the Red Branch (Fig.~\ref{fig:Plx_RB}) clearly indicates that the stars with redder $(BP-RP)$ colours present a mode well separated from that of the total sample (by $\approx$\  0.20 mas, attending to the position of the Kernel density maxima for both populations). Moreover, the RB FWHM is twice as narrow as that of stars of bluer colour. 

Our exploratory analysis thus suggests that the RB stellar population  shows distributions in the different planes that are compatible with, and representative of, a young and rich stellar cluster, from whose population we are just detecting the tip of the iceberg, i.e. the most luminous stars.  
We therefore conclude that the $(BP-RP)$ colour is an excellent  discriminator between cluster members and field stars for this stellar system. 

\subsection{Quantitative membership analysis: SALSON}
\label{sec:salson}

As a more sophisticated approach, we then proceeded to obtain individual membership probabilities by applying a non-parametric method based on the direct estimation of the probability density functions (pdf) by Kernel functions \citep[SALSON;][]{CCA1990}. This method is based on the assumption that cluster members are more densely concentrated than field stars in the space of variables. Eq.~\ref{eq:eq_pdf} defines the pdf of the class \textit{k} for the location of the object \textit{i}, in a variable space \textit{S}, using a Kernel, \textit{K}, where $\vec{S_i}$ represents the vectorial coordinates of the object \textit{i} in the space \textit{S}, as

\begin{equation}
D_{\vec{Si}}(k) =  \sum_{i\neq j}^{\forall\ j\  \in\ k\ \mathrm{class}} K_{S}(\Vec{S_i}, \Vec{S_j},  h_{S}) \, .
\label{eq:eq_pdf}
\end{equation}
 
We assume that there are only two classes (\textit{k}): cluster (\textit{c}) and  field (\textit{f}). We work with a Gaussian Kernel that only depends on the smoothing parameter \textit{$h_S$}. The space \textit{S} is formed by the four variables ($X,\ Y,\ \mu_{\alpha},\ \mu_{\delta}$), but considering that the subspaces $P=$\ ($X,\ Y$) and $V=$\ ($\mu_{\alpha},\  \mu_\delta$)  are statistically independent, in such a way that $D_{PVi}(k)$ can be estimated as $D_{PVi}(k)\ = \ D_{Pi}(k) \times D_{Vi}(k)$.
The separation of the space of variables allows us to determine three membership probabilities, one each corresponding to the subspaces \textit{P} and \textit{V}, both in 2D, and the third associated with the 4D space. These probabilities are defined by the expression

\begin{equation}
Pr_{\Vec{Si}}(k)
= \displaystyle \frac{Pr(k) \times D_{\Vec{Si}}(k)} {\sum_{\forall k} Pr(k) \times  D_{\Vec{Si}}
(k)} \, ,
\label{eq:eq_prob}
\end{equation}


\noindent where $Pr(k)$ 
are the \textit{a  priori} probabilities  that are estimated, in a frequentist inference approach, as the relative frequency of elements belonging to each class ($k$). Obtaining the pdfs, as well as the \textit{a posteriori}  probabilities, follows this procedure:
\begin{enumerate}
    \item Removing the most evident outliers in both subspaces.
    \item Estimating the Kernel smoothing parameters ($h_P$ and $h_V$) as those that maximise the likelihood in both subspaces ($P$ and $V$), respectively \citep{Silverman1986}. This requires performing the previous step; otherwise, the smoothing parameter estimates could be biased, and so could be the final probability assignment \citep{CCA1985}. 
    \item  Estimating the pdf in subspace $V$ for each object in the whole sample, and sort them from highest to lowest densities.
    \item Making the first classification between both classes, by assigning as cluster members the higher 0.4 percentile of the pdf in $V$, ($D_{Vi}$).
    \item Calculating $D_{Pi}(c)$, $D_{Pi}(f)$, $D_{Vi}(c)$, and $D_{Vi}(f)$, by using Eq.~\ref{eq:eq_pdf}. With these values, the probabilities $Pr_{Pi}(c)$, $Pr_{Vi}(c)$, and $Pr_{PVi}(c)$ are then estimated via Eq.~\ref{eq:eq_prob}.
  \item Classify as cluster membersing those objects with $Pr_{Vi}(c)\geq$ 0.5 \&\ $Pr_{PVi}(c)\geq$ 0.5. This criterion takes into account both subspaces, giving more weight to the kinematic variables.
  \item Comparing this classification with the previous one, and, if they do not coincide, returning to step 5. If they match, the run is stopped and an output file is written, containing the $D_{Si}(k)$ for the two subspaces and both classes, as well as the three estimated probabilities $Pr_{Pi}$, $Pr_{Vi}$, and $Pr_{PVi}$.
  \end{enumerate}

\begin{figure}
   \resizebox{\columnwidth}{!}
   {\includegraphics[angle=0, clip]{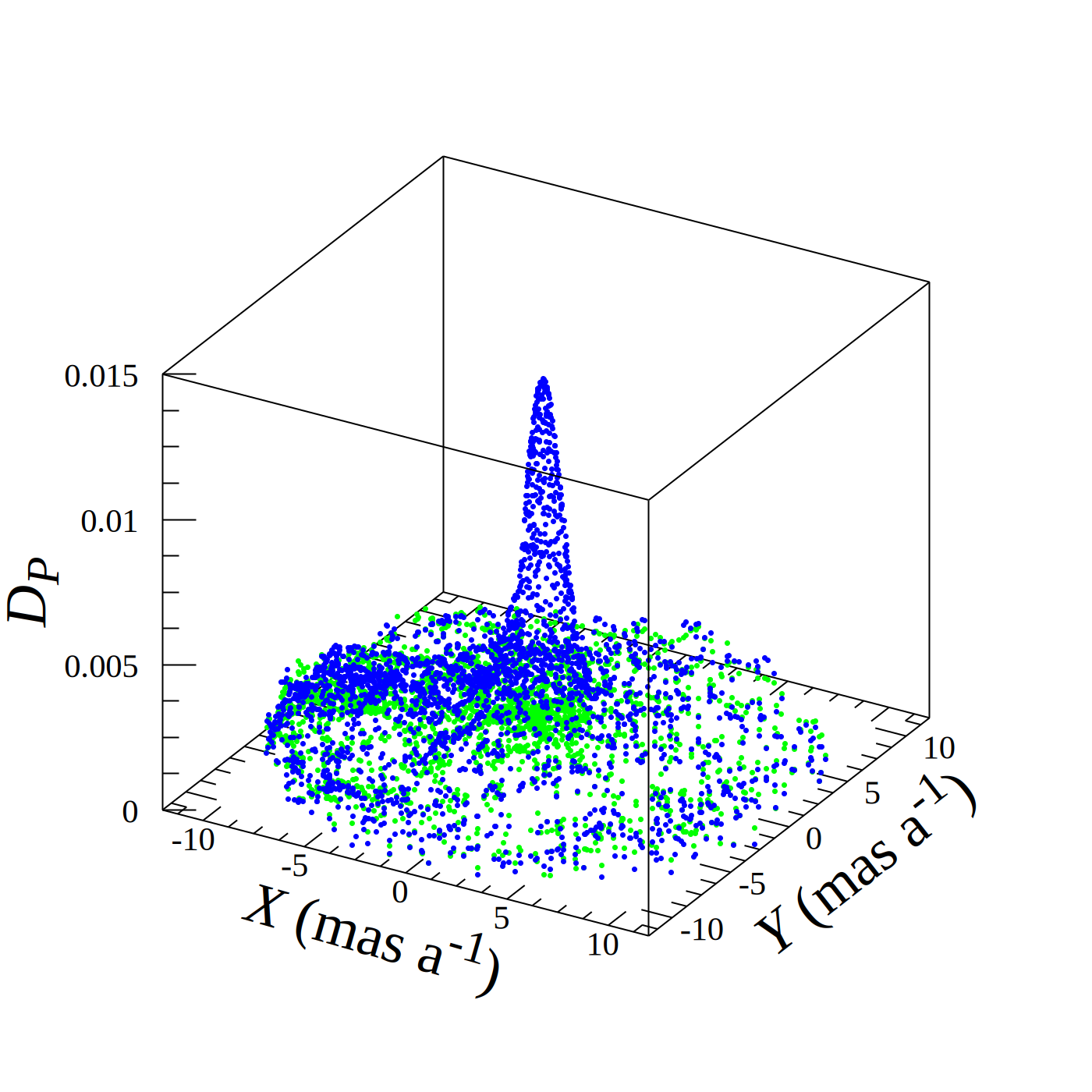}}
   \caption{$D_P$ pdfs on the X-Y plane for the probable clusters members selected by SALSON. $D_P(c)$ in blue and $D_P(f)$ in green. $D_P(c)$ and $D_P(f)$ show similar values for objects well outside the central core, which leads to larger uncertainties in the estimated cluster membership probabilities. 
}
   \label{fig:3D_members_P}
    \end{figure}

\begin{figure}
   \resizebox{\columnwidth}{!}
   {\includegraphics[angle=0, clip]{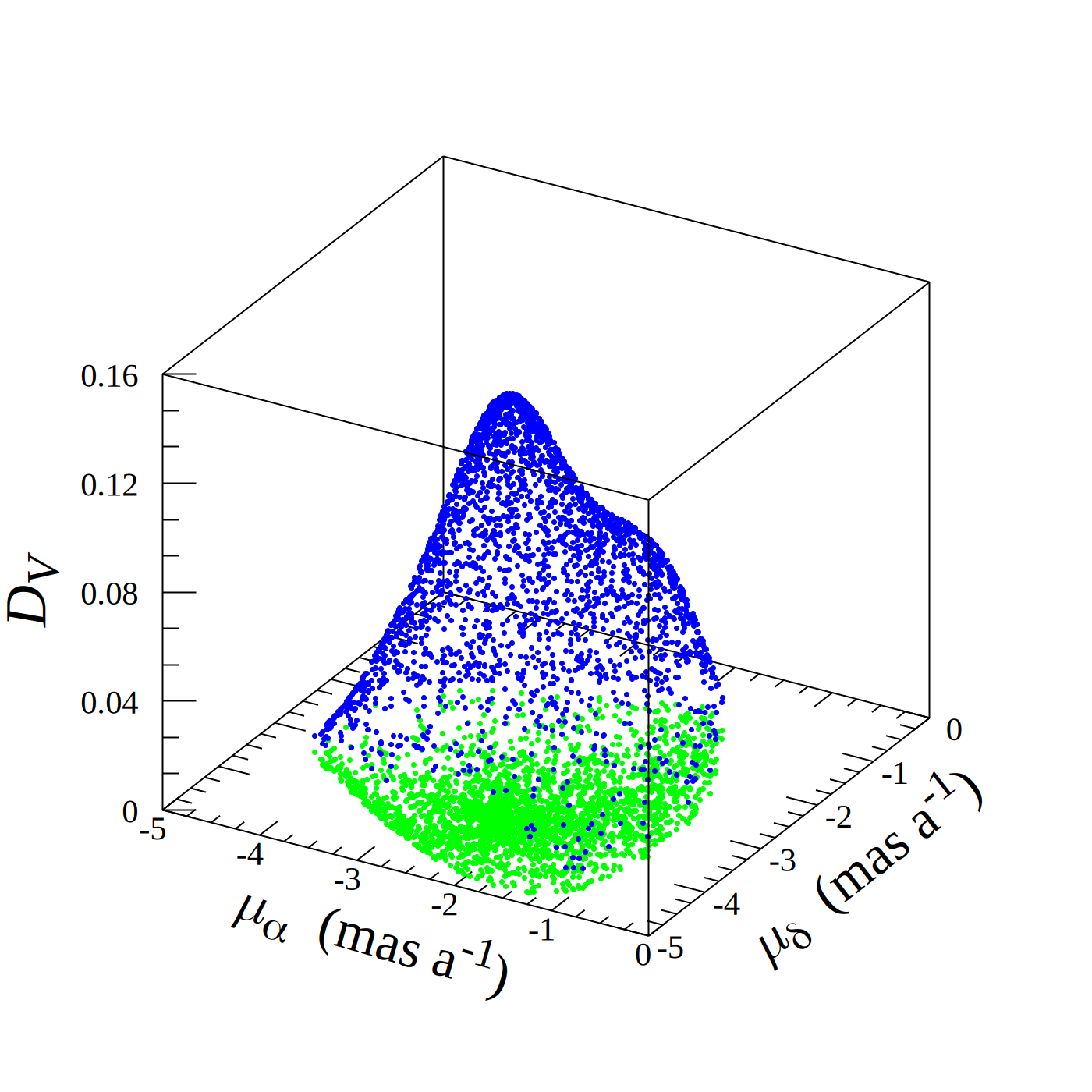}}
   \caption{$D_V$ pdfs in the proper motion space for the probable clusters members selected by SALSON. $D_V(c)$ is in blue and $D_V(f)$, in green. $D_V(c)$ shows a non-single peak distribution with a rounded bump at larger proper motions. This is likely indicative of a field contamination not removed by SALSON. Possible causes are discussed in the text.}
   \label{fig:3D_members_V}
\end{figure}

From the 19\,432 objects in our initial sample, SALSON identifies 8010 as cluster members. In Figures~\ref{fig:3D_members_P} and~\ref{fig:3D_members_V} we show the pdfs of the cluster class in subspaces $P$ and $V$. Stars identified as members distribute in positional space ($P$; see Fig.~\ref{fig:3D_members_P}) with two different shapes,  a central concentration  of stars,  within a raw external radius of around $3\farcm5$, and an extended halo, where the pdfs for the cluster ($D_P(c)$; blue) and field ($D_P(f)$; green) classes are much smaller and very close to each other, so that the probabilities given by Eq.~\ref{eq:eq_prob} can be very uncertain for these objects.

\begin{figure}
   \resizebox{\columnwidth}{!}
   {\includegraphics[angle=0, clip]{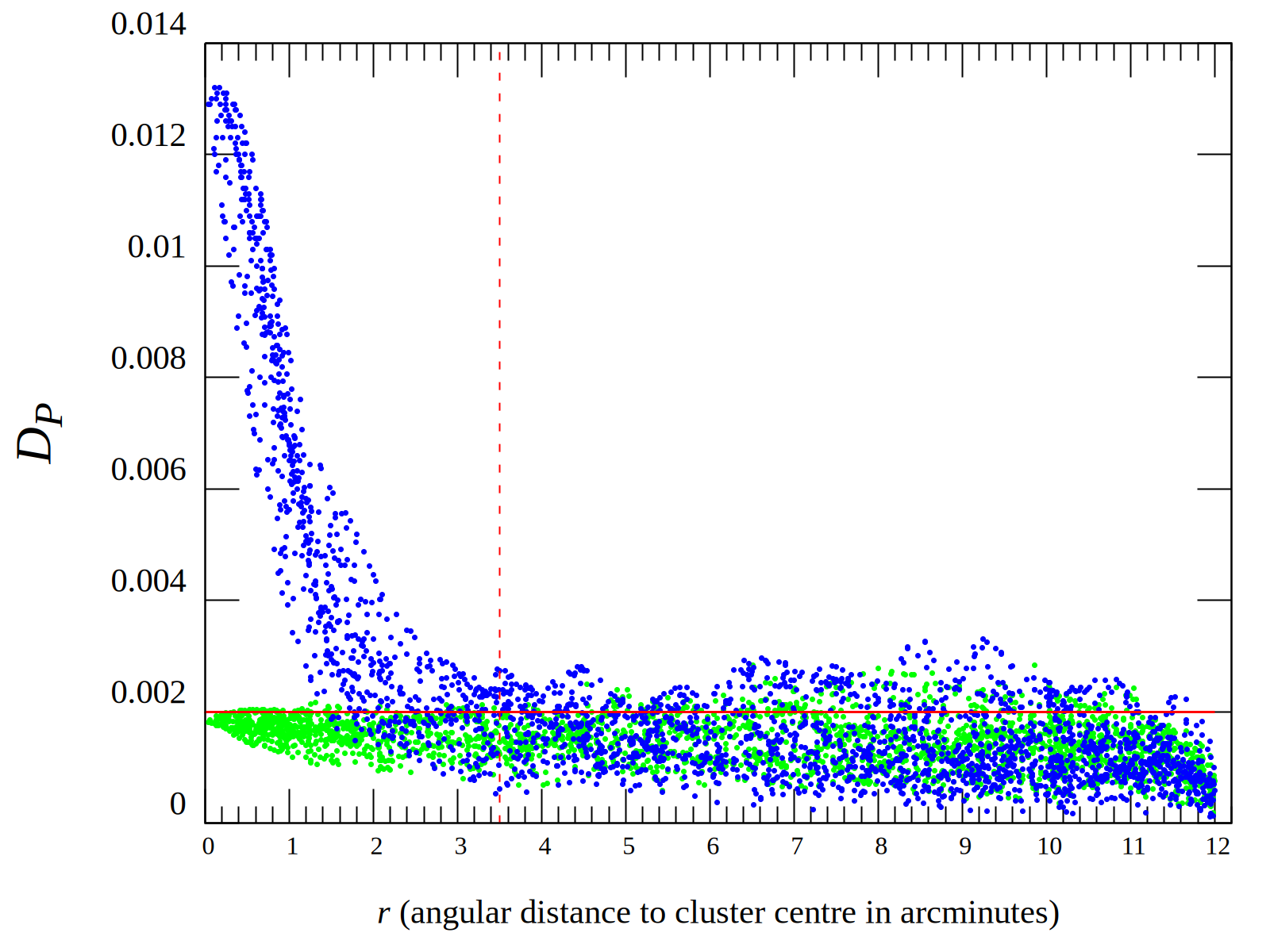}}
   \caption{Wd~1 radial density profile. The $D_P$ pdfs versus angular distance (arcmin) to cluster centre is shown for the members selected by SALSON. $D_P(c)$ is shown in blue and $D_P(f)$, in green. We identify as most likely cluster members those stars within the upper-left rectangle (red  lines): $D_P(c)\  >$\ 0.002 and $r \ <$\ 3.5 arcminutes. }
   \label{fig:DP_r}
\end{figure} 

In the proper motion subspace (Fig.~\ref{fig:3D_members_V}), there is a clear separation between both pdfs, defined by the plane $D_{V}(c) =$\ 0.02. The pdf in $V$ shows a rounded peak, but there is also a bump of stars adjacent to the main maximum, which suggests that we could 
have a non-negligible contamination from other stellar populations with different, but close, angular velocities. 

In any event, the number of objects selected is much higher than the number of objects with red colours that we identified in the previous section. Moreover, the stars selected as cluster members are distributed all over the photometric CMD. 
Therefore, we proceed to prune the dataset of those objects, which, despite their SALSON cluster member classification, show lower pdf values in both subspaces, as well as very similar values for both classes. To this aim, we plot the radial spatial distribution of the cluster by drawing the $D_{P}(c)$ (blue) and $D_{P}(f)$ (green) densities versus the radial angular distance to cluster centre for the objects selected as members (Fig.~\ref{fig:DP_r}). 

This plot suggests that most conspicuous members are within a radius of $\approx3\farcm5$ and show $D_{P}(c)$ larger than 0.002. Thus, stars enclosed in the upper-left rectangle of Fig.~\ref{fig:DP_r} are our best candidates according to their spatial distribution. By imposing this criterion plus that of the kinematic subspace ($D_{V}(c) >$\ 0.02), we extract 1154 high probability members supported by, and estimated from,  astrometric \textit{Gaia}~EDR3 data. Still, these objects display a variety of photometric colours and magnitudes that are incompatible with a single population. If we now apply the colour criterion that was identified in the previous section, 
($(BP-RP) >$ 4), we are left with a set of 401 \textit{bona fide} cluster  members that we later use as probes for obtaining accurate values of the physical parameters of Wd~1. We have to note that the selected cluster radius is larger than any other previous estimation, but we prefer to be conservative in this initial selection and come back later to a more refined calculation of the cluster radius. 

 \subsection{Wd~1 and its surroundings: Is there anything else?} 
 \label{subsec:colours}
 

In the previous section, we have selected a list of \textit{bona fide} cluster members by applying astrometric and photometric criteria. However, the simple application of a non-parametric method to the astrometric variables separates a much larger population, prompting two questions: Is there a second kinematic population hidden in this selection? Is our application of a colour cut fully justified?

To answer these questions, we proceed to analyse the spatial distribution of the $(BP-RP)$ colour for the 8010 probable cluster members selected by SALSON. The goal is to find whether there are narrow ranges of colour that are spatially concentrated in the RA-DEC subspace. For this, we use the \textit{Spectrum of Kinematic Groupings} algorithm \citep[SKG;][]{AlfaroGonzalez2016,AlfaroRoman2018}  based on the \textit{Minimum Spanning Tree} (MST) which was first proposed for the analysis of mass segregation \citep{Allison2009, Maschberger2011}, but has been easily extended to other physical variables that may also be spatially segregated \citep{GonzalezAlfaro2017, CostadoCygnus2017, CostadoMonoceros2018, Carballo2021}.

The basic foundations of the method rely on sorting  the variable to be analysed (ascending or descending order does not matter), selecting the number of points per bin, which is tentatively taken as the integer closer to the square root of the number of total points, and choosing  another integer that stands for the number of overlapping points between two consecutive intervals. The algorithm estimates the MST of each bin in the position space and the median of the distances between adjacent points of the MST. This value is compared with that obtained for a sample of the same size extracted from the original sample (full range of the variable). If the quotient between this last median and that of the corresponding bin is significantly greater than 1, we say that this interval of the variable is spatially segregated, or, in other words, that condition
 \begin{equation}
   \Lambda_i - (2\times \sigma_{\Lambda_i}) >\ 1  
   \label{eq:Lambda_sigma}
 \end{equation}
 \noindent is fulfilled. For a more detailed explanation on how $\Lambda_i$ and $\sigma_{\Lambda_i}$ are estimated, we refer the reader to \cite{AlfaroGonzalez2016}. A more conservative criterion can be chosen  by simply increasing the multiplicative factor of $\sigma_{\Lambda_i}$ in Eq.~\ref{eq:Lambda_sigma}.
The latest implementation of the algorithm \citep{AlfaroRoman2018}  also calculates the value of $Q$ \citep{Cartwright2004} as a quantitative description of the internal spatial structure of the objects in each bin. Values above 0.8 indicate central distributions. Clumpy patterns present values below 0.8 and a homogeneous distribution would show values around 0.8 \citep{Cartwright2004, SanchezAlfaro2009} . The precision in the estimation of $Q$ is obtained by bootstrapping \citep{AlfaroRoman2018}.

\begin{figure}
   \resizebox{\columnwidth}{!}
   {\includegraphics[angle=0, clip]{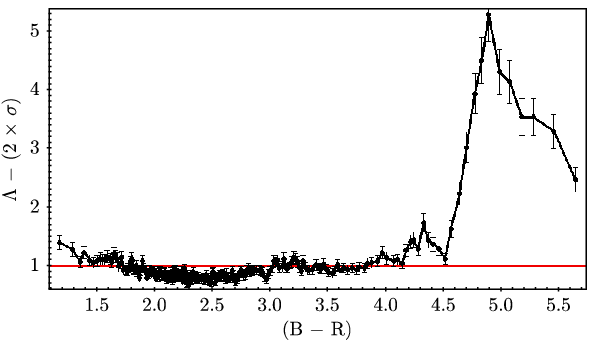}}
   \caption{Spectrum of the spatial segregation by $(BP-RP)$ colour. Error bars are $\sigma$ values. $\Lambda$\, –\, $(2\times\sigma)$ values higher than 1 are indicative of colour bins spatially grouped. A strong spatial segregation is observed for stars redder than $(BP-RP)\, >$\, 4. A blue grouping is also detected with $(BP-RP)\, <$\, 1.5.
}
   \label{fig:SKG_BR}
\end{figure}

\begin{figure}
   \resizebox{\columnwidth}{!}
   {\includegraphics[angle=0, clip]{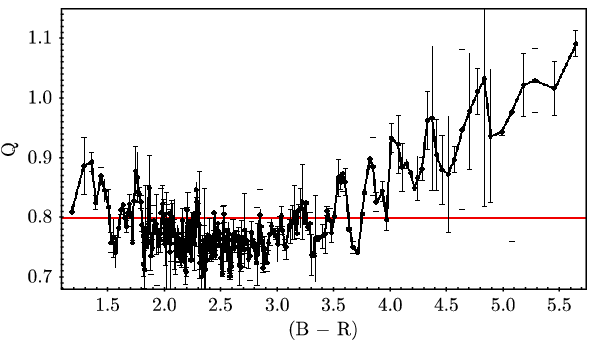}}
   \caption{$Q$ spatial concentration parameter as a function of colour. Error bars have been obtained by bootstrapping. $Q \approx$\ 0.8 separates the clumpy and central spatial distributions. The higher $Q$ is, the more spatially concentrated the distributions. This plot reinforces the conclusions derived from Fig.~\ref{fig:SKG_BR}}
   \label{fig:Q_BR}
\end{figure}

By applying this technique to the 8010 probable members outlined by SALSON, we obtain the SKG shown in Fig.~\ref{fig:SKG_BR}. Along the $x$ axis we plot the central $(BP-RP)$ colour of each bin, while on the vertical axis we display $\Lambda_i - ( 2\ \times\  \sigma_{\Lambda_i})$. The plot clearly highlights the increase of $\Lambda_i$ with growing colour (starting around $(BP-RP) \approx$ 4), which indicates a sharp spatial segregation for the redder stars, peaking at  $(BP-RP) \approx$ 5\footnote{Although we do not use this information for the analysis at this point, in Section~\ref{sec:discuss}, we see that this is approximately the colour of the bulk of OB members (see Fig.~\ref{fig:cmd_members}).}. Fig.~\ref{fig:Q_BR} shows the plot of $Q$\  vs\  $(BP-RP)$ colour. Again, the reddest stars show a significant increasing trend of $Q$ with growing colour that indicates that they are more centrally concentrated the redder they are, in agreement with the results observed in the SKG (Fig. \ref{fig:SKG_BR}). On the blue side of the colour interval ($ (BP-RP) <$ 1.5), a spatial segregation with a central concentration ($Q >0.8)$ is also observed, although with $\Lambda$ and $Q$ values much lower than those shown by the red tip of the colour interval. Based on this result, we then select two data sets of spatially segregated stars: a) the red group (RG) formed by stars with $(BP-RP)>$ 4, and b) the blue group (BG) formed by objects with $(BP-RP)< 1.5$. Both groups present well differentiated distributions of the astrometric variables in \textit{Gaia}~EDR3. 

\begin{figure}
   \resizebox{\columnwidth}{!}
   {\includegraphics[angle=0, clip]{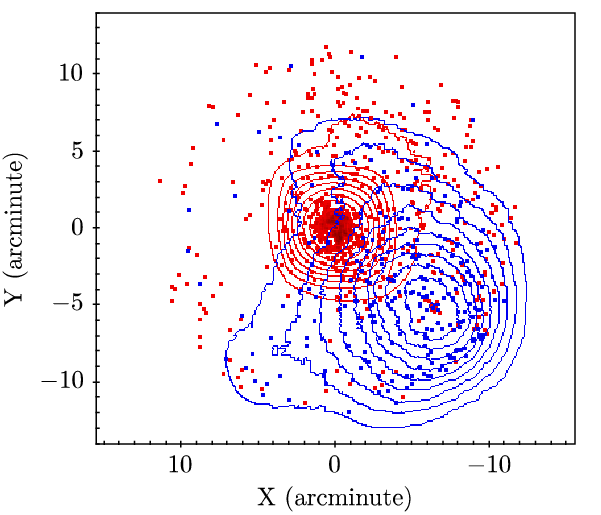}}
   \caption{Stars (dots) and isodensity contours (lines) for the two  groups spatially segregated by $(BP-RP)$.  BG (in blue) contains the objects in segregated bins with colour bluer than 1.5. RG (in red) is the same for colours redder than 4. The RG is mainly associated with the cluster centre, while the BG shows a stellar population with its maximum density well separated from the core of Wd~1. The axes show angular distances in equatorial coordinates.}
   \label{fig:Grupos_lb}
\end{figure}

\begin{figure}
   \resizebox{\columnwidth}{!}
   {\includegraphics[angle=0, clip]{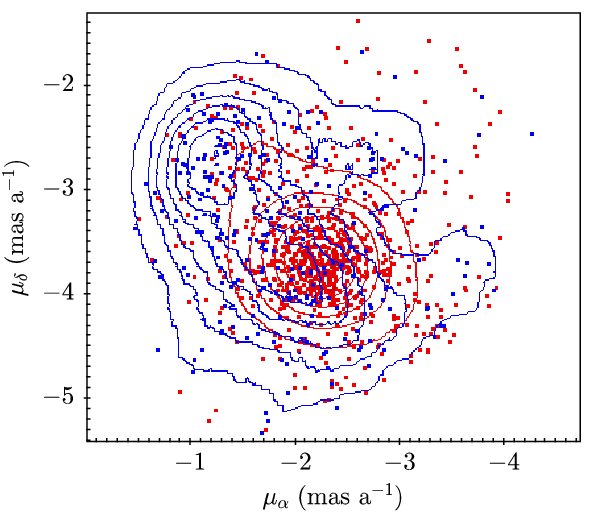}}
   \caption{Stars and isodensity contours in proper motion space  for the two  groups spatially segregated by $(BP-RP)$. Symbols and colours as in Fig.~\ref{fig:Grupos_lb}. The spatial segregation by colour is mimicked in the VPD. Although both distributions show a non-null overlap, their density maxima are well separated in proper motion. 
}
   \label{fig:Grupos_pm}
\end{figure}

\begin{figure}
   \resizebox{\columnwidth}{!}
   {\includegraphics[angle=0, clip]{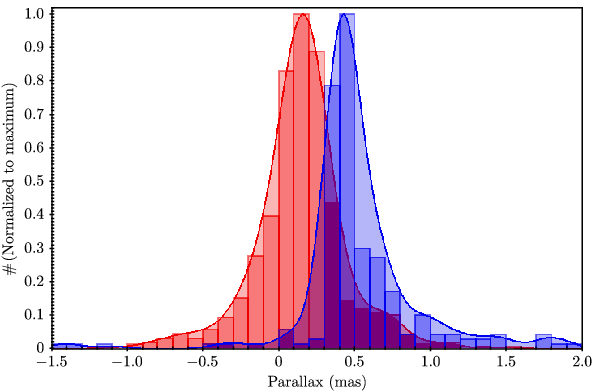}}
   \caption{Histogram of corrected (see Section~\ref{subsec:correct} for parallax bias corrections) \textit{Gaia}~EDR3 parallaxes for the RG (red) and BG (blue) groups, showing two different statistical distributions. RG shows smaller parallaxes than BG. This result, together with those shown in Figures~\ref{fig:SKG_BR} to~\ref{fig:Grupos_Plx}, strongly suggests that both populations are well separated by colour, position, kinematics, and distance. Gaussian Kernel density estimators are also drawn.}
   \label{fig:Grupos_Plx}
\end{figure}

To visualise this, we show the distribution of both groups in equatorial coordinates (Fig.~\ref{fig:Grupos_lb}), proper motions (Fig.~\ref{fig:Grupos_pm}) and parallaxes (Fig.~\ref{fig:Grupos_Plx}). The isodensity contours for both groups have also been drawn for  descriptive purposes in the position and proper motion diagrams. The RG is mainly associated with the known population of Wd~1, as already discussed in previous sections. The BG appears to be distributed over the area located to the south-west of the cluster. The  density local maximum is  around ($l$, $b$) = ($339\fdg41$, $-0\fdg38$), but the population is much less concentrated. Similarly, the RG is densely concentrated in the \textit{Vector Point Diagram}, while the BG seems to be  associated with a second bump of lower height, well separated from the central value of cluster proper motion. However, the most conclusive evidence that we are observing two different stellar systems is provided by the histogram of the parallaxes (Fig.~\ref{fig:Grupos_Plx}) which shows two clearly different distributions. We performed a two-sample Kolmogorov-Smirnov test for the parallaxes of the groups RG (771) and BG (222). The statistic $KS$ is 8.198, which leads to a probability $p <$ 0.00001, whose value is much lower than 0.05, therefore rejecting the hypothesis that the two samples come from the same distribution. It is clear then that the BG represents a grouping located closer to the Sun, with a central parallax (mode) implying a raw distance of around 2~kpc. 

The distribution of the RG and BG groups on the plane of the sky appears to be primarily drawn by the interstellar extinction pattern in the field. Both groups extend mainly over the western region of the area analysed (Fig.~\ref{fig:Grupos_lb}), while to the east of the cluster the surface density of objects decreases significantly. Going into more detail, while the RG is distributed over most of the western semicircle, the BG is concentrated towards the south-west of the cluster.

The nature of the BG group is unclear. We discuss it in Appendix~\ref{app:bh197}. Independently of this, the analysis presented in this section leaves two conclusions: 1) the simple application of a non-parametric method to the astrometric variables (positions and proper motions) does not separate effectively the cluster population from its surroundings, and 2) the population with redder colours appears as clearly distinct in both physical space and astrometric parameter space, fully justifying our choice of colour as a discriminant.

\section{Distance estimation and kinematic parameters}
\label{sec:distance}

Once we have selected a safe set of cluster members, we can use their  \textit{Gaia}~EDR3 parallaxes to estimate the distance to the cluster. Unfortunately, the quality of astrometric data for \textit{bona fide} cluster members is low, because of crowding and faintness, and we cannot calculate bias corrections for a majority.

\subsection{Parallax bias correction}
\label{subsec:correct}

Parallaxes catalogued in \textit{Gaia}~EDR3 present systematic biases (or zero points) that depend on several factors, such as the photometry of the star, its location on the sky, ecliptic latitude and  number of  orbits, as well as the set of free parameters used in fitting the astrometric solution \citep{Lindegren2021,maiz22}. Those two references provide independent estimates of the systematic biases that are quite similar for faint stars ($G > 13$) but are somewhat different for brighter objects. The biases are of the order of a few tens of mas and can be corrected by employing the Python algorithm provided by the first set of authors\footnote{
\url{https://gitlab.com/icc-ub/public/gaiadr3\_zeropoint}} and by the IDL algorithm provided by the second author as an appendix to his paper (which can be used for both zero points). From now on, we refer to the first systematic bias as the Lindegren zero point (Lztp) and to the second one as the Ma\'{\i}z~Apell\'aniz zero point (MAztp).

We applied both algorithms to the whole sample of \num{19432} objects. A total of \num{12070} stars have zero-point estimates, which represents 62\% of the initial sample. This drastic reduction in the number of objects is mainly due to the position of Wd~1 in a very crowded region, showing a wide range of magnitudes and colours, but also because we have only recovered interpolated solutions (see \citealt{Lindegren2021}, and the corresponding tutorial\footnote{\url{https://gitlab.com/icc\-ub/public/gaiadr3\_zeropoint}.}, for a more detailed explanation of these issues). In other words, we do not work with extrapolated zero points for the sake of accuracy, at the cost of reducing the final number of cluster members. The effect on the \textit{bona fide} members is much worse than on the whole sample, as we are dealing with some of the most crowded and reddest stars. From 401 members selected in Section~\ref{sec:salson}, only 142 stars have interpolated solutions. Of those, 138 have five-parameter astrometric solutions and four have six-parameter astrometric solutions.

\begin{figure}
   \resizebox{\columnwidth}{!}
   {\includegraphics[angle=0, clip]{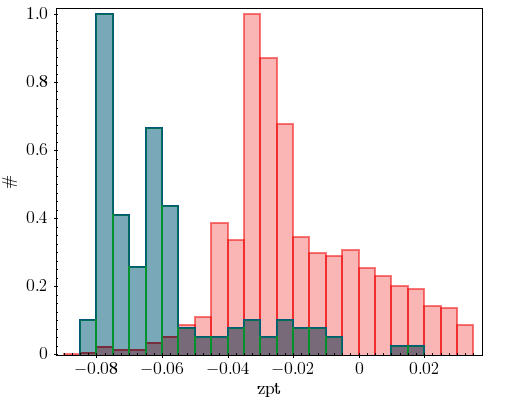}}
   \caption{Histogram of parallax zero-point corrections according to \citet{Lindegren2021}. Bona fide members are in blue, and the global sample is in red. The histograms have been normalized to their maxima for better visualizing the different distribution modes. }
   \label{fig:hist_zpt}
\end{figure} 


\begin{figure*}
   \centerline{
   \includegraphics[width=0.49\textwidth]{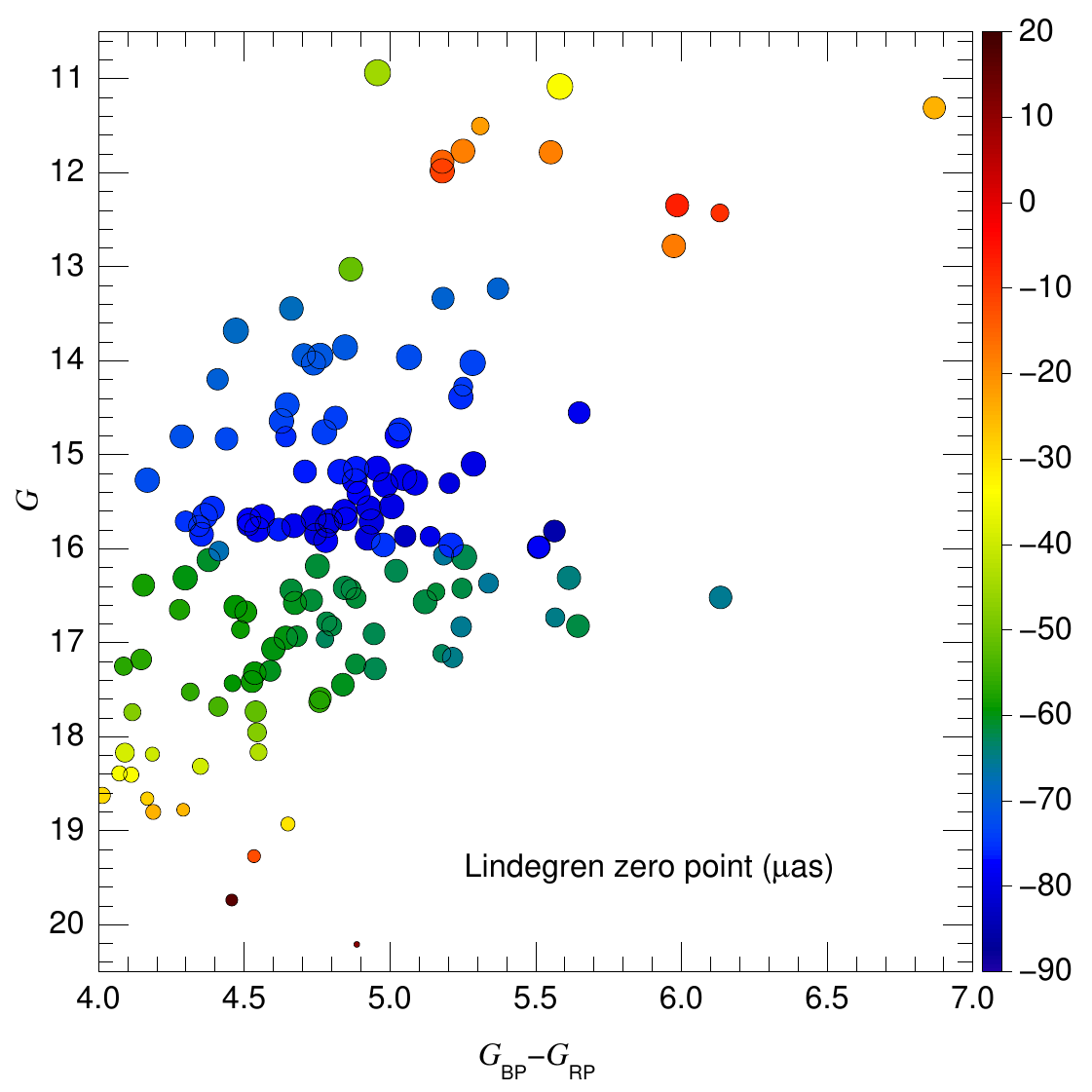} \
   \includegraphics[width=0.49\textwidth]{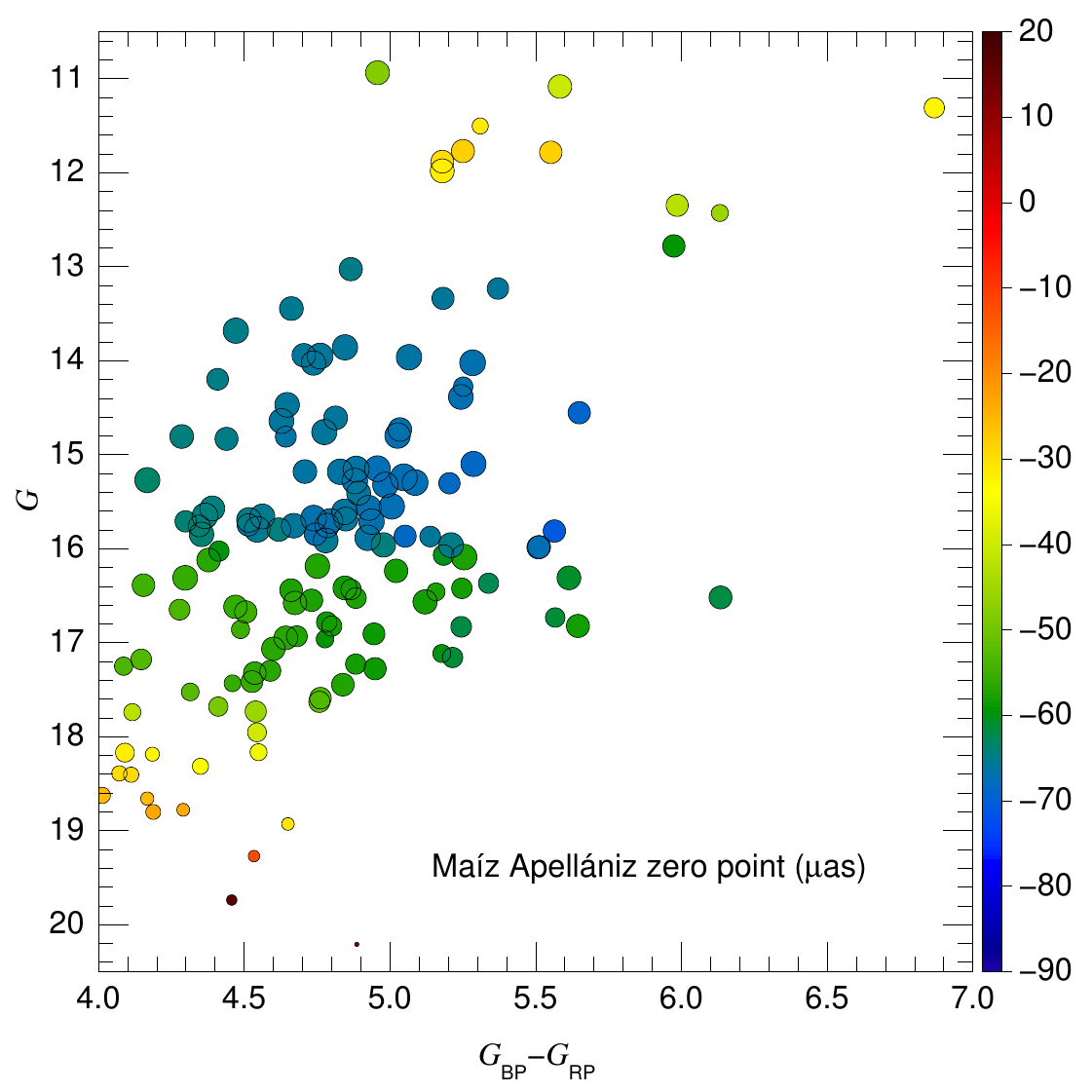} 
   }
   \caption{Cluster CMD showing the value of the zero point applied to each one of the \textit{bona fide} members of Wd~1 for which such correction has been estimated. (left panel) Lindegren. (right panel) Ma\'{\i}z~Apell\'aniz. The scale is in $\mu$as and larger symbols have larger full weights.}
   \label{fig:zpt_GBR}
\end{figure*} 

\subsection{Cluster parallax}
\label{sec:parallax}

Once we have corrected the individual parallaxes from the zero point biases estimated in the previous section, we proceed to determine the cluster parallax and its uncertainty. 
Until now, we have not filtered the data by astrometric errors, either in proper motions or parallaxes; we only applied a RUWE threshold. Now, we do introduce the errors catalogued in \textit{Gaia}~EDR3 to estimate the total uncertainty of the parallax and use them as a weight parameter (alone or together with other parameters connected to the cluster membership) in determining the mean parallax of the cluster and its error. To this goal, we follow the procedure proposed by \citet[][and references therein]{maiz22}.

Parallax errors catalogued in \textit{Gaia}~EDR3 
data release do not include the actual uncertainty in the measurement of that variable. However, the true value  can be approximated \citep[see discussion in][]{maiz21a} by the expression:

\begin{equation}
 \sigma_{\varpi_{\mathrm{e}}}=\sqrt{k^{2}\sigma^{2}_{\varpi,i} + \sigma^{2}_{\varpi,\mathrm{s}}}  \, .
\label{eq:true_var}
\end{equation}

The values involved in this prescription can be different for different releases and the subscripts $i$, s, and e indicate formal, systematic, and total errors, respectively. The multiplicative constant $k$ depends on several factors, and there are different empirical approaches for its estimation. We consider two possibilities: (a) the estimate from \citet{maiz21a} when applying Lzpt 
and  (b) the estimate from \citet{maiz22} when applying MAzpt. 

Both approaches (and others, such as \citealt{Vasiliev2021}, as well) yield similar values for $k$, which do not differ significantly from 1.3 for most of the stars in our sample with high weights. For $\sigma_{\varpi,s}$, we use the value of 10.3~$\mu$as from \citet{maiz21a}.

In Fig.~\ref{fig:hist_zpt} we show the histograms of the Lzpt 
correction for the entire sample (red) and for the \textit{bona fide} members (blue). Both histograms have been normalized to their respective maximum to better distinguish distribution modes. While the majority of stars have corrections of $\approx-0.03$~mas, close to the average value for \textit{Gaia}~EDR3 given by \cite{Lindegren2021}, cluster members typically have corrections between $-0.05$ and $-0.085$, with the mode at $-0.08$~mas. 
The photometric values ($G$ and $(BP-RP)$) are the main variables that influence the high absolute value of this correction for cluster stars, as can be seen in Fig.~\ref{fig:zpt_GBR}.

The value of the cluster unbiased parallax can be estimated by the expression
\begin{equation}
    \varpi_{c} =\sum_{i=1}^{n}w_i \varpi_{i}\, ,
\label{eq:unb_plx}
\end{equation}

\begin{table*}[]
\caption{Parallax and distance results using different assumptions.}
\centering
\renewcommand{\arraystretch}{1.3}
\begin{tabular}{cccccccc}
\hline
case & solutions    & zero point            & astrometric & weights  & $N$ & $\varpi_c$ & $d$                    \\
     &              &                       & parameters  &          &     & ($\mu$as)  & (kpc)                  \\
\hline
A    & all          & MAzpt & 5+6         & parallax & 401 & 234$\pm$12 & $4.30^{+0.24}_{-0.22}$ \\
B    & interpolated & MAzpt & 5+6         & parallax & 142 & 243$\pm$13 & $4.14^{+0.24}_{-0.21}$ \\
C    & interpolated & MAzpt & 5+6         & full     & 142 & 238$\pm$12 & $4.23^{+0.23}_{-0.21}$ \\
D    & interpolated & MAzpt & 5           & parallax & 138 & 243$\pm$13 & $4.14^{+0.24}_{-0.21}$ \\
E    & all          & Lzpt             & 5+6         & parallax & 401 & 233$\pm$12 & $4.32^{+0.24}_{-0.22}$ \\
F    & interpolated & Lzpt             & 5+6         & parallax & 142 & 247$\pm$13 & $4.07^{+0.23}_{-0.21}$ \\
G    & interpolated & Lzpt             & 5+6         & full     & 142 & 242$\pm$12 & $4.16^{+0.22}_{-0.20}$ \\
H    & interpolated & Lzpt             & 5           & parallax & 138 & 247$\pm$13 & $4.07^{+0.23}_{-0.21}$ \\
\hline
\end{tabular}
\renewcommand{\arraystretch}{1.0}
\label{parallax_results}
\end{table*}

\noindent where the weights $w_{i}$ can be defined in two different ways by assigning different weights: a) by taking into account only the true uncertainty of the parallaxes, as in \citet{maiz21a}, or b) also including information related to the membership analysis that we have carried out in Section~\ref{sec:salson}. In this way, two different aspects are involved in the weighted average, the parallax error, which is intrinsic to the star and its measurement, and the congruence of its positional and kinematic values within the cluster, which somehow measures its connection to the stellar system itself. Both conditions constrain and refine the final cluster parallax. The expressions that we use are:
\begin{equation}
w_{i}=\displaystyle{\frac{1/\sigma^{2}_{e,i}}{\sum_{i=1}^{n}1/\sigma^{2}_{e,i}}}
\label{eq:weight_1}
\end{equation}
\noindent and
\begin{equation}
 w_{t,i} = \displaystyle\frac {(1/\sigma^{2}_{e,i})\times D_{P,i}(c)\times D_{V,i}(c)}{\sum_{i=1}^{n}(1/\sigma^{2}_{e,i})\times D_{P,i}(c)\times D_{V,i}(c)} 
 \label{eq:weight_3}
\end{equation}

\noindent for the case when only parallax error is taken into account (a; Eq.~\ref{eq:weight_1}), and for the case when we also include the pdfs for the cluster class in both subspaces (positional and kinematic) in the computation (b; Eq~\ref{eq:weight_3}). We refer to Eq.~\ref{eq:weight_1} as the parallax weights and to Eq~\ref{eq:weight_3} as the full weights.

The error of the weighted mean parallax is obtained with the formula
\begin{equation}
\sigma^{2}_{\varpi_{c}}=\sum_{i=1}^{n}w_{i}^{2}+2\sum_{i=1}^{n-1}\sum_{j=i+1}^{n}w_{i}w_{j}V_{\varpi}(\theta_{ij}) \:\,\:.
\label{eq:covariance}
\end{equation}
\noindent Here, the second term on the right side of the equation represents the angular covariance, where  $V_{\varpi}(\theta_{ij})$ is given by equation~8 in \citet{maiz21a}. This analytic function depends on the covariance angular value, $V_{\varpi}(0)$, at the limit  $\theta \xrightarrow{}$ 0,  which  is taken as 100 $\mu$as  in this work.

We present the results for the parallax to Westerlund~1 in Table~\ref{parallax_results}. We show eight different combinations, by varying the type of solution used (all or only interpolated), the zero point applied (MAzpt or Lzpt) 
the number of astrometric parameters for the stars (five and six or just five), and the weights used (parallax or full). In each case, the number of stars in each sample, the cluster parallax, and the cluster distance are given. For the distances we use the prior of \citet{Maiz01a,Maiz05c}, which is specific for OB stars, with the parameters of \citet{Maizetal08a}.

The results for the eight cases in Table~~\ref{parallax_results} are very similar, with the two members of any given pair within one sigma or less of each other. This indicates that our distances are robust and  independent of the selection we choose from the sample of probable members,  
or which zero point is used. From now on, we adopt the case C distance of $4.23^{+0.23}_{-0.21}$~kpc.


\subsection{Cluster kinematics}
\label{subsec:otherpar}

Once the sample of \textit{bona fide} members of Wd~1 has been selected and its distance estimated with the procedure detailed in the previous section, we now proceed to determine the main kinematic variables of the cluster. For this, we now use the whole sample of 401 cluster members. 


As with the parallaxes, the proper motions in \textit{Gaia}~EDR3 show systematic errors that, although negligible in most cases, need to be corrected  if we want to determine the motion of the cluster in the Galaxy with adequate precision to analyse its current and future dynamic state. The zero points for proper motion corrections have been determined following the procedure described in \cite{CG2021}. The estimates depend mainly on the celestial positions and on the $G$ magnitude of the stars. In the case of the 401 stars in our sample, this correction is limited to an average zero point of $-0.02$~mas/a  in $\mu_\delta$. Given the cluster coordinates,  and the stellar brightness range, the zero point in $\mu_\alpha$ is negligible. The mean values and variances of the $\mu_\alpha$ and $\mu_\delta$ components  have been estimated following the same procedure as for the determination of  parallax. A value of $\mu = [-2.231 \pm\ 0.008, -3.697 \pm\ 0.008]$, in Equatorial coordinates, has been obtained in mas/a units by using the full weights from the previous Subsection.



\section{Spectroscopic data}
\label{sec:spectra}

\subsection{Description of the sample}

\begin{figure*}
   \resizebox{\columnwidth}{!}{\includegraphics[angle=0, clip]{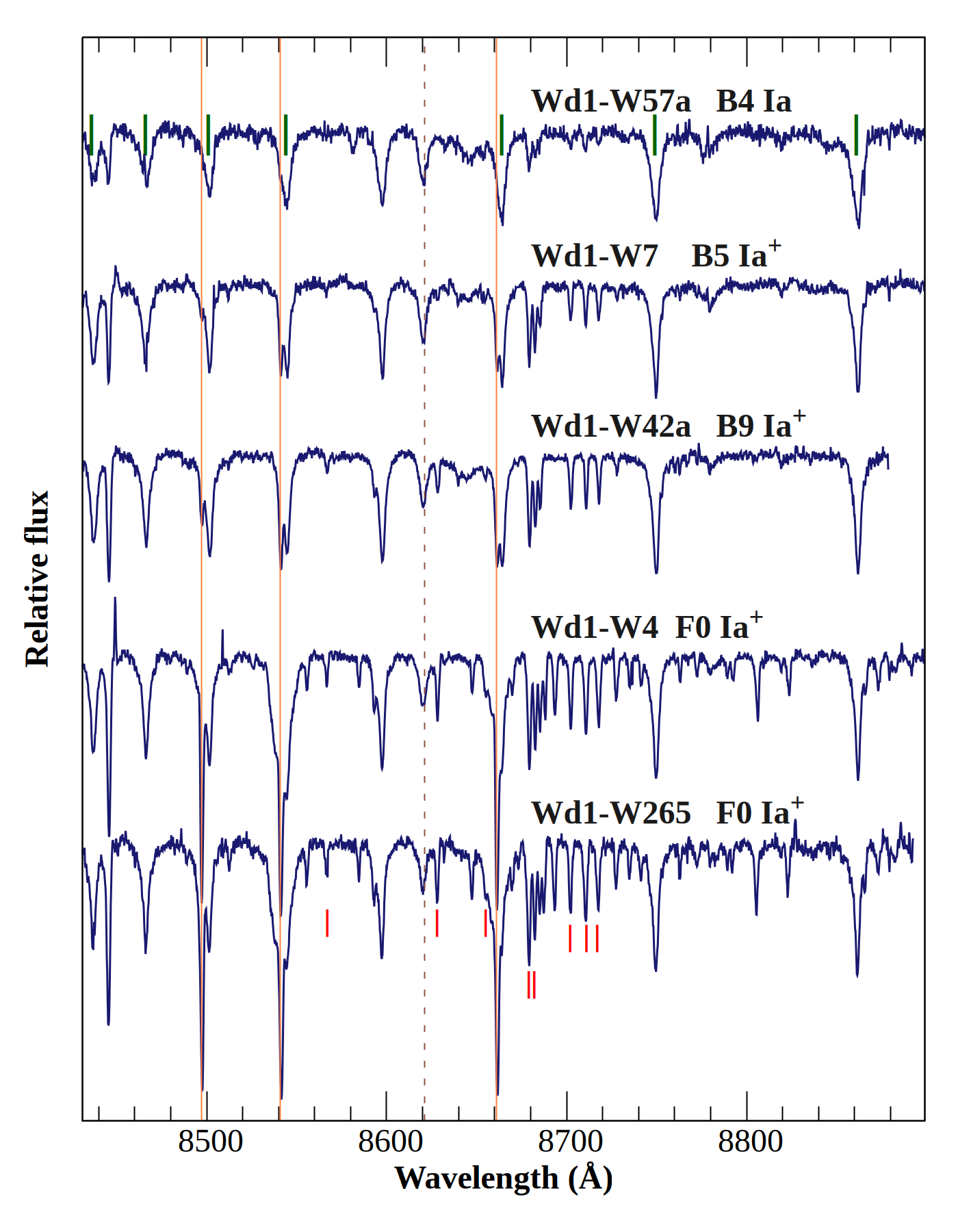}}
   \resizebox{\columnwidth}{!}{\includegraphics[angle=0, clip]{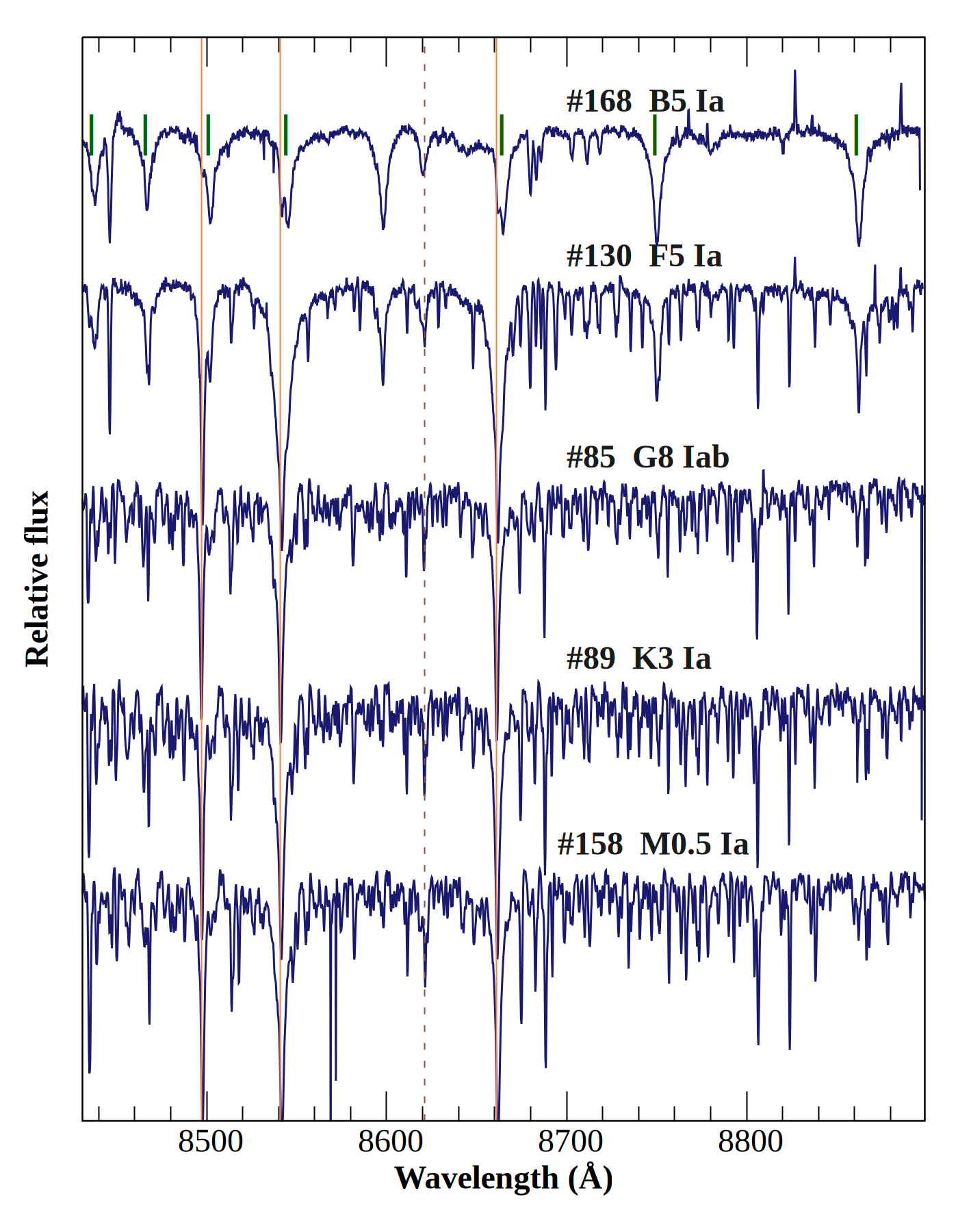}}
   \caption{AAOmega spectra of luminous supergiants. All the spectra have been normalised and the continuum has been shifted for clarity. \textit{Left panel:}  Members of Wd~1. The (green) dashes above the spectrum of Wd1-W57a indicate the positions of Paschen lines. The three thin (orange) vertical lines show the position of the \ion{Ca}{ii} triplet. The short (red) dashes below the spectrum of Wd1-W265 identify some of the strongest \ion{N}{i} lines. The dotted line marks the position of the 8621\,\AA\ DIB, whose strength correlates well with extinction up to a saturation value. \textit{Right panel:} Other stars in the field, not connected to the cluster. Symbols as in the left panel.}
   \label{brightomega}
\end{figure*}

In total, our AAOmega observations include 14 early-type members of Westerlund~1, the LBV Wd1-W243, two yellow hypergiants, Wd1-W4 and Wd1-W265, and the four red supergiants. Spectra of the most luminous stars are displayed in Fig.~\ref{brightomega} (left panel). The stars that were observed are indicated in the penultimate column of Table~\ref{clarkmembers}. Five of the early-type stars are located outside the well studied core region (of radius around $2\arcmin$). Four of them were subsequently observed with FLAMES and are thus included in the catalogue of spectroscopic members in \citet{clark20}, as W1049, W1053, W1067, and W1069. The fifth is more than $5\arcmin$ away from the centre, but its proper motions are within 1-$\sigma$ of the cluster average and its position in the CMD confirms it as the most distant known member. Following the naming convention in \citet{clark20}, we identify this object (Gaia EDR3 \num{5940104594038054144}) as Wd1-1070. It is further discussed in Appendix~\ref{app:members}.

Outside this central concentration, which corresponds to the cluster, there are 29 other "blue" targets, none of them closer than $20\arcmin$ from the cluster centre. Of these, four objects are catalogued early-type stars with usable blue spectra. All other sources, with only $I$-band spectra, are new identifications, except for star \#~119 = [GKF2010]~MN48, which has subsequently been identified as an LBV \citep{kniazev16}, and is further discussed in Appendix~\ref{app:omega}. The list of targets observed is shown in Table~\ref{tab:2dFblue}. Spectral types have been estimated following the criteria discussed in \citet{negueruela10Wd1}. See Appendix~\ref{app:omega} for further details.

In addition, we have spectra for 321 "red" targets. Of them, only two are within $5\arcmin$ of the cluster centre. As is typical of the criteria used, a large fraction of the stars observed turn out to be late-M giants, in all likelihood AGB stars, whose colours are indistinguishable from those of red supergiants. The list of targets observed and their characteristics can be found in Table~\ref{tab:2dFred}. Spectral types have been estimated following the criteria discussed in \citet{negueruela12}. We find around 100 stars whose spectral features allow a supergiant classification (luminosity class Ib-II or above). Of these, close to 40 can be unambiguously identified as luminous red supergiants (RSGs), i.e. massive stars.


Despite a moderate success rate among RSG candidates (fully within the expectations of the method), between the \textit{blue} and \textit{red} targets, we identify at least 70 previously unknown massive stars within $1\degr$ of Wd~1. A representative sample of good quality spectra is shown in Fig.~\ref{brightomega} (right panel), where the whole range of spectral types is represented.

\subsection{Radial velocities}

For the majority of the stars in the ``red" sample (spectral types G and later), it was possible to calculate radial velocities by means of the method described in detail in \citet{gonzalez2015}. In short, we cross-correlated each observed spectrum against the most similar MARCS synthetic spectrum in the grid available from the POLLUX database \citep{pal2010}. In \citet{gonzalez2015}, this method resulted in a typical uncertainty (velocity dispersion for stars observed more than once during that run) around $1.0~\mathrm{km\,s^{-1}}$. As that sample was observed with the same instrumental configuration and on the same dates as our data, we can assume with confidence the same uncertainty for our measurements. Finally, we calculated the heliocentric and Local Standard of Rest corrections through the spectroscopic-analysis software \textit{ispec} \citep{ispec2014,ispec2019}. 
 
For 115 objects, we can compare the RVs measured from our spectra to those measured by \textit{Gaia}~DR2. The average difference (our heliocentric velocities minus those from DR2) is $-0.7\:\mathrm{km}\,\mathrm{s}^{-1}$ (below our  uncertainty of $1.0~\mathrm{km\,s^{-1}}$), with a standard deviation of $3.0\:\mathrm{km}\,\mathrm{s}^{-1}$. There are only five stars whose differences are $>6\:\mathrm{km}\,\mathrm{s}^{-1}$ in modulus. These values suggest that our RVs are in exactly the same reference system as those from \textit{Gaia}~DR2, and we can safely use our much larger sample.

Unfortunately, the vast majority of our targets are faint and have very poor astrometric solutions, and any attempt to derive individual distances results in very large uncertainties. For completeness, in Table~\ref{tab:2dFred}, we give the distances derived by \citet{bj21}. If we select only objects that we have classified as RSGs and impose a quality criterion $\varpi/\sigma_{\varpi} > 6$, there might be a trend to more negative radial velocities with increasing distance, as expected from the radial velocity curve in this direction. However, no object with a \citet{bj21} distance $>3 \: \mathrm{kpc}$ passes the quality criterion, and thus our objects cannot be used to explore the radial velocity distribution up to the distance of Westerlund~1 with the current EDR3 dataset.

\section{Discussion}
\label{sec:discuss}

We have investigated the population of massive stars in the field surrounding Westerlund~1 and the kinematic properties of the cluster and its surroundings. The first conclusion that may be drawn from the previous analysis is that this is a very complex field. A direct application of non-parametric Bayesian inference (SALSON) to separate the cluster from the field results in a sample with an important field contamination. 
The stars identified as cluster members, based \textit{only} on sky positions and proper motions, represent 40\% of the initial sample, but are spread over the parameter space both in parallax and in the CMD. This field contamination is mainly caused by four reasons: 1) the cluster proper motions are quite similar to those of many foreground stars along the line of sight (see Fig.~\ref{fig:VPD_RB}), 2) the actual cluster members represent a very small fraction of the total sample,  3) almost all actual members have large errors in their astrometric parameters (see Table~\ref{clarkmembers}), precluding their identification as a distinct population, and 4) the large sampling radius, much larger than the cluster size, results in a higher pollution by field stars \citep{SVA2010, sanchez20}, which we finally remove by restricting the sample to the central $3\farcm5$.

In spite of this, the spatial distribution of members selected by SALSON is strongly concentrated towards the position of Westerlund~1. There are two main reasons for this: 1) actual cluster members are very heavily concentrated, driving the behaviour of the whole subsample, and 2) Westerlund~1 is seen through a hole in the extinction, so that the density of foreground stars is also much higher in its vicinity. In fact, the density of stars in the SALSON sample is high over the whole south-western quadrant, including the cluster itself, and much lower in the other three quadrants.

\begin{figure}
   \resizebox{\columnwidth}{!}
   {\includegraphics[angle=0, clip]{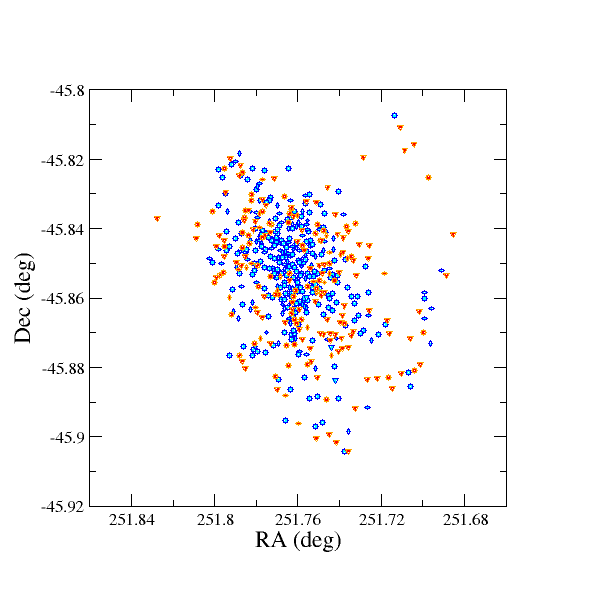}}
   \caption{Spatial distribution of \textit{Gaia} EDR3 \textit{bona fide} (circles) and likely (defined as objects picked by SALSON as possible members and having $3.5\leq BP-RP<4.0$; triangles) members of Westerlund~1, divided by brightness. Blue symbols represent stars brighter than $G=17$, while red symbols are fainter objects. There are only 5 bright likely members. See text for details.}
   \label{fig:member_distribution}
\end{figure} 

Given this high contamination, some additional information is needed to allow a better selection of cluster members.
For this, we resort to the $(BP-RP)$ colour, after our analysis shows that 1) redder stars are much more strongly concentrated on the sky and in the vector point diagram than the rest of the sample and 2) all known members are among the reddest stars in the field. We note, however, that -- unlike \citet[]{beasor21}
-- we do not start from the known members and we do not limit our analysis to stars with similar proper motions. As our primary aim is determining the distance to the cluster, we do not use parallax information in the selection of cluster members. We allow the non-parametric analysis of proper motion and spatial distribution, together with a colour cut, to select the cluster members.  Nevertheless, our sample includes most of the members listed in \citet[]{clark20}. As can be seen in Table~\ref{clarkmembers}, most of those not included have poor astrometric parameters, thus confirming the validity of our approach. Without the additional information provided by the spatial distribution and colour cuts, the sample would be dominated by Scutum Arm stars whose proper motions are moderately similar to those of cluster members, while their astrometric errors are much smaller. These circumstances very likely explain the results of \citet[]{aghakan21}. 

To evaluate how effective our member selection is, we consider the catalogued members brighter than $RP=13.5$, a range for which \citet{clark20} are likely complete and where we can expect astrometric parameters to present reasonable errors. There are 39 of such objects in \citet[][]{clark20}. Our algorithm identifies 27 as cluster members. Of the 12 objects not identified as \textit{bona fide} members, one (Wd1-W9) has no astrometric solution in EDR3, one (WR~T) is outside the $r=3\farcm5$ threshold and eight have RUWE $>1.4$. The only two objects with good astrometry that are not considered \textit{bona fide} members by the algorithm are Wd1-1049 and Wd1-1067, two halo members whose proper motions are compatible with the cluster average, but that just fail to pass the cut in positional probability. 

Conversely, essentially all the astrometric \textit{bona fide} members are part of the cluster population. We find only two interlopers among the $>40$ astrometric members brighter than $G=15.5$ (all the other ones have counterparts in the catalogue of \citealt{clark20}). The first one is \textit{Gaia} EDR3 \num{5940106208947388416}, which has colours similar to the RSGs in the optical, but is much fainter in the near-infrared. We have a spectrum of this object (our target \#238) which shows it to be a very late red giant. The second one is \textit{Gaia} EDR3 \num{5940106625576004992}, which we identify with object F1 in \citet{ritchie09bins}, a field red giant that happens to have proper motions compatible with membership. Therefore, we can safely assume that most of our 401 astrometric members are OB cluster members.

Nevertheless, our sample of \textit{bona fide} members is very far from complete. There are some bright members with poor astrometry, among which we can cite the RSG Wd1-W75, or the luminous supergiants Wd1-W28, Wd1-41 and Wd1-43b. But the effects of crowding become much more important for fainter stars. To illustrate this point, in Fig.~\ref{fig:member_distribution}, we plot the spatial distribution of our members, divided into a bright ($G<17$) and a faint ($G\geq 17$) sample. Almost all the stars in the faint sample come from the periphery of the cluster. There is not a single faint member selected in the densest regions, demonstrating how complicated this field is for \textit{Gaia}. This can be easily visualised by looking at Table~\ref{clarkmembers}, which contains the EDR3 parameters for the cluster members listed in \citet{clark20}. Stars with $G\approx17$ that lie in crowded regions are lacking parameters (e.g. Wd1-W6b, Wd1-W14c, WR~J, WR~K) or have large errors in their proper motions that prevent their identification as members (e.g. WR~G, WR~H), while others of similar brightness in less crowded regions have smaller errors and are picked up by SALSON (e.g. WR~Q, or even Wd1-1064, with $G=17.6$, and WR~W with $G=18.2$).

It could be argued that our colour cut at $(BP-RP)=4.0$ is to some degree arbitrary, and a slightly bluer or redder colour could have been chosen. The analysis presented in Section~\ref{subsec:colours} suggests that cluster members start to be numerous around $(BP-RP)=3.5$. There are a few catalogued members with $(BP-RP)$ just below 4~mag (we can cite Wd1-1010, Wd1-1013, Wd1-1020 and Wd1-W228b). They are all located in the south-western tip of the cluster, clearly the area of the lowest extinction. Moving our threshold to $(BP-RP)>3.5$ adds about one hundred extra likely members. Given their spatial distribution, heavily concentrated towards the cluster, their membership is very likely. To illustrate this, they have been added to Fig.~\ref{fig:member_distribution}, where they appear as triangles. Interestingly, they are all faint. Except for the four objects mentioned above and one other uncatalogued source, they are all fainter than $G=17$. The absence of any bright stars at lower reddening indicates that stars with ($BP-RP$)\,$<4$ are a minor component of the cluster population. Like their more reddened siblings, these faint objects form a halo around the cluster core. Again, no faint astrometric members are detected in the densest regions, strongly hinting that crowding is preventing the detection of faint stars in the cluster core. The degree of incompleteness in the magnitude range $G=17$\,--\,19 is very difficult to evaluate, as the quality of \textit{Gaia} data (whether a star has an astrometric solution, useful uncertainties or even $BP$ and $RP$ magnitudes) depends strongly on the object's local environment.

In any event, the main source of bias in our sample is the faintness of stars in the $BP$ band, due to the enormous reddening. The CMD for our \textit{bona fide} members (Fig.~\ref{fig:cmd_members}) clearly shows that we are missing a very large number of stars because they are too faint in $BP$. This is the only sensible explanation for the diagonal edge to the cluster sequence on its red side, while the $(J-K_{\mathrm{S}})$ vs
$K_{\mathrm{S}}$ CMD is essentially a vertical strip extending over six magnitudes \citep[e.g.][]{gennaro2017}. As a rough guide to the level of incompleteness, we cross-matched the list of \textit{bona fide} members displaying $G<17$ with the photometry of \citet{gennaro2017}. We then counted the number of objects lying in the same region of the $(J-K_{\mathrm{S}})$ vs
$K_{\mathrm{S}}$ CMD as these stars. Taking into account that the photometry of \citet{gennaro2017} is also incomplete (due to both crowding and saturation in the vicinity of the brightest members), we estimate that the \textit{Gaia} sample is $>50\%$ incomplete in this luminosity range.

\subsection{The size of Wd~1}

In optical images, Wd~1 has a distinctive shape, resembling a crescent, with two well separate groups, the main one to the north and a smaller aggregate to the south. Contrarily, our selection of members is distributed in an elliptical shape, with a rather smooth distribution, as also found by \citet{gennaro2017}, based on near-infrared star counts. These two disparate observations need to be reconciled.

\begin{figure*}
   \resizebox{\textwidth}{!}
   {\includegraphics[angle=0, clip]{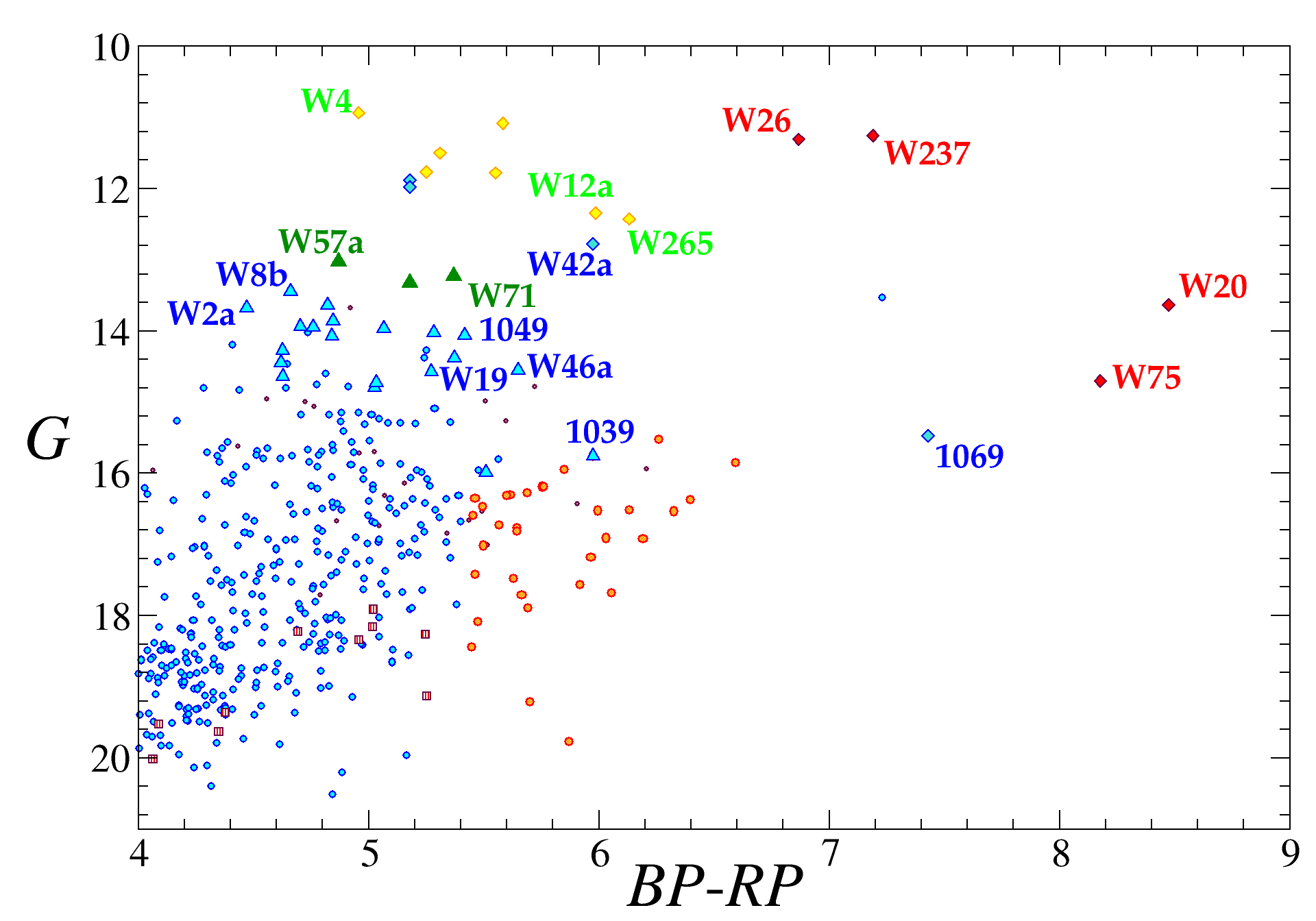}}
   \caption{Cluster CMD for all the \textit{Gaia} EDR3 \textit{bona fide} members of Westerlund~1 and spectroscopic members from \citet{clark20} not selected by the algorithm. Diamonds represent the most luminous hyper- or supergiants (according to colour, blue, yellow -- including the LBV Wd1-W243 -- and red). Triangles are luminous B-type supergiants (those classified as Ia). Green triangles have mid-B types, while turquoise triangles have early-B types. The bulk of OB members is represented by circles. The larger blue circles represent the astrometric members, while the smaller brown circles are stars in \citet{clark20} not selected by the algorithm. The orange circles represent the candidate high-reddening sample. The red squares are objects that present much redder infrared colour than typical, $(J-K_{\mathrm{S}})>2.2$.}
   \label{fig:cmd_members}
\end{figure*} 

\begin{figure}
   \resizebox{\columnwidth}{!}
   {\includegraphics[angle=0, clip]{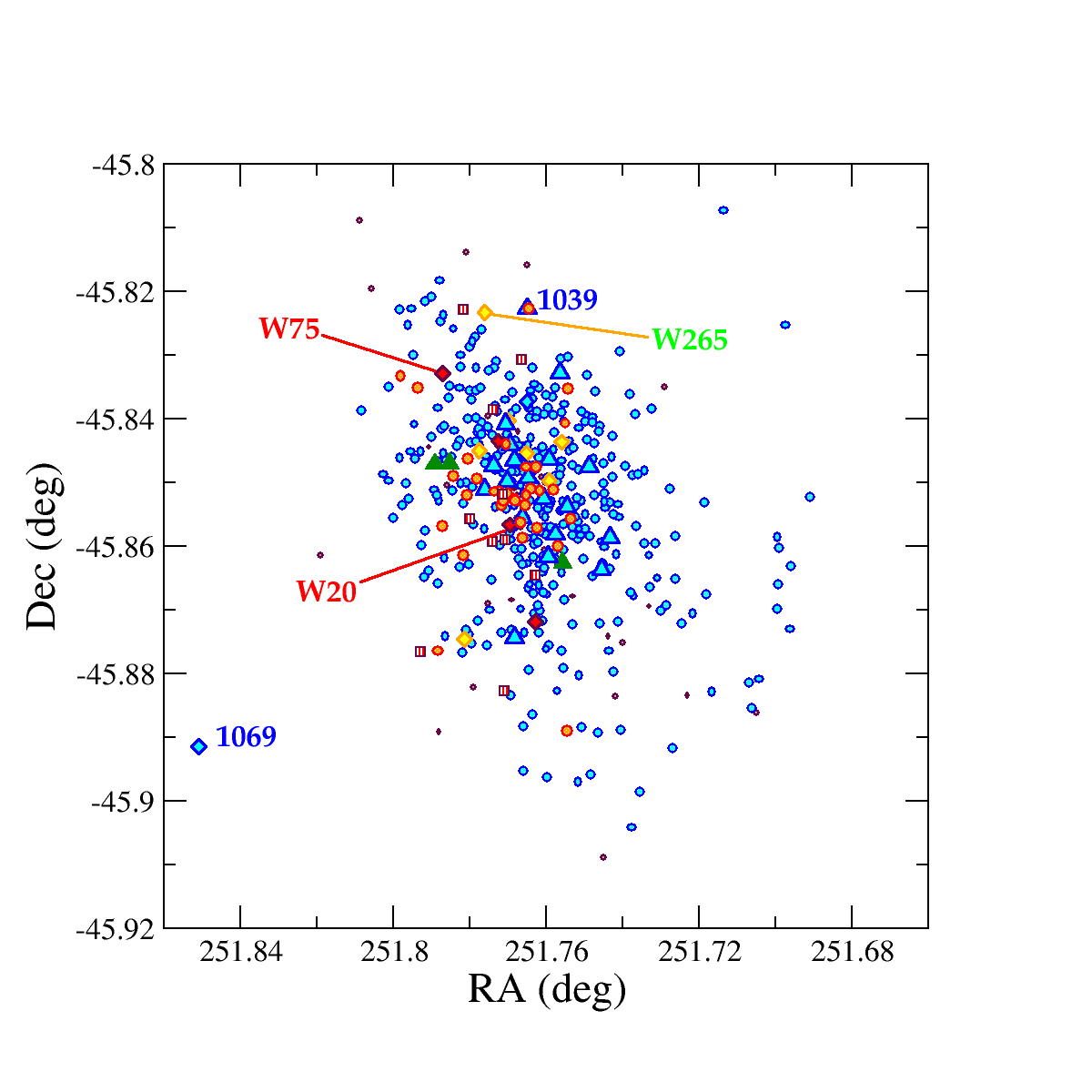}}
   \caption{Spatial distribution for the objects shown in Fig.~\ref{fig:cmd_members} (symbols as in that figure). The spatial concentration of the orange circles, presumably high-reddening OB stars, stands out. A few objects of interest discussed in the text are labelled. Wd1-1039 is the only object with two labels, as it is both a Ia supergiant and a highly reddened star (Wd1-1069 is much more heavily reddened, but is not marked with an orange circle, as it lies outside the main distribution). A handful of the members in \citet[][]{clark20} are outside the area shown.}
   \label{fig:pos_members}
\end{figure}

When we look at the CMD for \textit{bona fide} members (Fig.~\ref{fig:cmd_members}), we find a very large spread of colours. Leaving aside the strip of cool hypergiants at the top, we find a significant number of objects whose $(BP-RP)$ colour is much higher than the average or mean\footnote{This value includes the colours of the RSGs and YHGs, which are intrinsically much redder than the bulk of the population, made up of OB stars.} for the members (4.7 and 4.8~mag, respectively). To understand this distribution, we selected a sample of candidate OB stars with high reddening, taking the red edge of the cluster sequence, at $(BP-RP)\ga5.4$. We note that this edge may not truly represent the reddest cluster members, as it is likely determined by the observability of objects in the $BP$ band. When we look at the spatial location of these objects (Fig.~\ref{fig:pos_members}), we find that they are very tightly concentrated: most of them are distributed in a narrow strip between the two main concentrations mentioned in the previous paragraph or immediately to the east of this strip. Such concentration immediately suggests that these objects represent a population with higher extinction.

To confirm this point, we cross-matched our selection of \textit{bona fide} members with the near-infrared photometry of \citet{gennaro2017}. When this photometry is plotted, these candidate high-reddening stars display values of $(J-K_{\mathrm{S}})\approx1.8$\,--\,2.0, as opposed to the 1.5\,--\,1.6 shown by the bulk of the cluster members. This difference of $\approx 0.3$~mag in $E(J-K_{\mathrm{S}})$ corresponds, by using the extinction law of \citet{damineli16}, to an extra extinction of $A_{V}\ga 2$~mag. Given their position in Fig.~\ref{fig:cmd_members}, this additional reddening implies that several of these stars have intrinsic magnitudes comparable to those of the luminous OB supergiants at the top of the cluster sequence. This is confirmed by their $K_{\mathrm{S}}<10$ magnitudes.  Despite this, most of them were not identified as candidate members by \citet[]{clark05} or subsequent works, presumably because they are not detectable in their $UBV$ photometry. They have never been observed spectroscopically, and thus they were not used by \citet[]{damineli16} to calculate the average extinction to the cluster. As an example, in Fig.~\ref{fig:cmd_members}, we identify the position of Wd1-1039, which is classified as B1\,Ia. Unfortunately, this object has no infrared photometry in \citet{gennaro2017}.

Interestingly, there is a significant number of stars that show even redder infrared colours, with $(J-K_{\mathrm{S}})\approx2.3$. They are also concentrated to the east of the cluster, although not so strongly. Unexpectedly, some of them have \textit{Gaia} counterparts, which are faint objects of moderately red colours (marked as stripped squares in Fig.~\ref{fig:cmd_members} and Fig.~\ref{fig:pos_members}). Their nature is unclear. They may be objects with intrinsic infrared excess, such as Herbig Be stars\footnote{Non-dusty Wolf-Rayet stars have similar colours, but the survey by \citet[]{crowther06WRs} should have detected all such objects, as it used near-infrared photometry.}. Alternatively, given their very faint $BP\approx22$ magnitudes and large photometric errors, their optical colours might be unreliable. Spectroscopic observations will be needed to explore these possibilities.

In any event, there is a substantial population of \textit{Gaia} members that have redder colours than the bulk of the population, indicative of higher extinction. Their spatial distribution leads us to conclude that the appearance of Wd~1 in optical images is driven by differential extinction. The range of extinction values present in the cluster is broader than previously assumed, with many infrared-bright sources simply not visible in the blue. Moreover, since \citet[]{damineli16} used only objects with spectral classification from \citet[]{negueruela10Wd1} to calculate individual reddenings and extinctions, their sample is strongly biased to low values. The true average extinction across the face of the cluster must be higher than their estimate. In this respect, it is worth noting that \citet{andersen17} found a significantly higher value for the average cluster extinction and a higher dispersion than \citet{damineli16} when using a technique based on individual dereddening of a sample of PMS stars observed with HST\footnote{Although the difference with respect to \citet{damineli16} may also be related to the extinction law chosen, \citet{andersen17} found that the extinction was higher to the north and north-east of the cluster, a conclusion borne out by the very red colours of objects such as Wd1-W75 and Wd1-W265 in Fig.~\ref{fig:cmd_members}. Unfortunately, they could not apply this technique to the central regions of the cluster because of saturation caused by the brightest stars.}.

This realisation has strong implications for our knowledge of the cluster basic properties. On the one hand, the combination of \textit{Gaia} and near-IR colours suggest that about 15 intrinsically luminous OB supergiants had not been previously identified. This compares to the approximately 55 OB supergiants of luminosity class Ia or Iab listed in \citet[]{clark20}, resulting in an increase of over 25\% in the cluster post-MS population.

Moreover, the assumption used by \citet[]{beasor21} to calculate bolometric luminosities for red and yellow hypergiants of an almost homogenous extinction for the cluster is shown not to hold. In fact, the RSGs Wd1-W20 and Wd1-W75, which these authors find to be of lower luminosity (implying an older population), are very evidently much more reddened than most of the cluster members, as can be seen in Fig.~\ref{fig:cmd_members}. Their near-IR colours, $(J-K_{\mathrm{S}})\approx3.8$ and 3.6, respectively, are about a whole magnitude redder than those of the two other cluster RSGs\footnote{The spectral types of all these objects are similar, with the possible exception of Wd1-W75, which is around M0\,Ia in the observations reported here, i.e. significantly earlier than the rest despite the redder colours. The other RSGs have been shown to be variable in spectral type \citep[]{clark10}, a feature typical of the most luminous M supergiants, but the \textit{Gaia} colours represent long-term averages.}. Wd1-W20 is surrounded by some of the most heavily reddened OB stars in the new sample (see Fig.~\ref{fig:pos_members}). Likewise, from their position in Fig.~\ref{fig:cmd_members}, the YHGs W12a and W265, which \citet[]{beasor21} find to be of low luminosity, are obviously more reddened than other stars of similar types. In contrast, Wd1-W4 or Wd1-W26, which according to \citet[]{beasor21} are much more luminous, happen to be rather less extinguished.


In fact, given the bias against stars with faint $BP$ magnitudes discussed above, we might be missing cluster members with very high reddening. In support of this, we mention the case of the BHG Wd1-1069. This star, which was first located with the observations reported here (our target \#355), and later repeatedly targeted during FLAMES monitoring (see \citealt{clark20}; their fig.~3), lies about $4\arcmin$ to the south-east of the cluster centre (see Fig.~\ref{fig:pos_members} for reference). Since it is outside the radius of $3\farcm5$ imposed to select \textit{bona fide} members, it is not in our list. Its proper motions are consistent with those of the cluster within their errors. As these uncertainties are large, we cannot decide if it is a halo member or a slow ejection from the cluster. Wd1-1069 is an extremely bright blue hypergiant, which displays $(J-K_{{\mathrm S}})=2.2$ and an extreme $(BP-RP)= 7.4$ (see its position in Fig.~\ref{fig:cmd_members}), which implies an absolute magnitude around $M_{K_{\mathrm{S}}}=-8$ at the cluster distance, similarly to its spectral twins Wd1-W7 and Wd1-W33. With $BP=21.1$, this object is at the limit of detectability with \textit{Gaia}, suggesting that any objects fainter in the blue (essentially any, except for other blue hypergiants) and affected by the same amount of reddening will not be detected by \textit{Gaia} in the blue band.

\subsection{The sightline to Wd~1}

The {\sc stilism} tool\footnote{At \url{https://stilism.obspm.fr/}.} \citep{capitanio17} indicates that in this general direction ($l=339\fdg5$, $b=-0\fdg4$) extinction is low at short distances, reaching on average only $E(B-V)\approx0.3$ at $1\:$kpc and $\approx0.5\pm0.1$ at $2\:$kpc. Extinction is, however, highly patchy, as the bright O9\,Ib supergiant HD~\num{151018}, which is projected exactly on top of the cluster and has a \textit{Gaia}~EDR3 distance of $2.05^{+0.08}_{-0.06}\:$kpc \citep{bj21}, suffers $A_{V}=2.9$ \citep{maiz18}. 

\begin{figure}
   \resizebox{\columnwidth}{!}
   {\includegraphics[angle=0, clip]{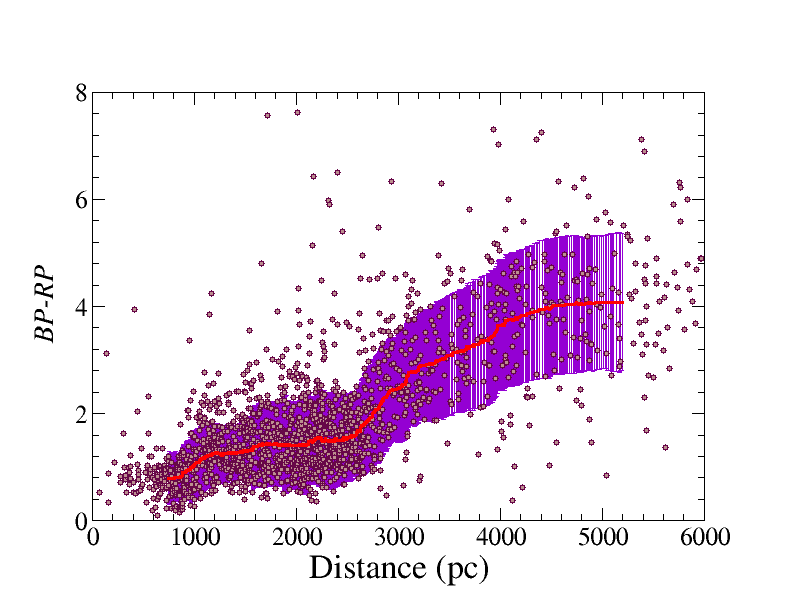}}
   \caption{Observed colour of stars in the field likely to be of early type (as estimated from their 2MASS colours) against their \textit{Gaia}~EDR3 distances \citep{bj21}. The thick (red) line is a 200-point running average. There is an obvious decrease in the number of objects around $d\approx2.8\:$kpc, which coincides with a steep increase in the average colour. The shaded area represents the running standard deviation. See the text for the many biases present.}
   \label{extinction_run}
\end{figure}

To investigate how extinction grows in this direction, we implement the following procedure: from the 2MASS catalogue, we take only sources with high precision photometry, by requiring all three magnitudes to have photometric errors $<0.03$~mag. We then select objects with $-0.1<Q_{\mathrm{IR}}<+0.1$, corresponding to early-type stars\footnote{This selection assumes a "standard" extinction law, different from that found by \citet{damineli16} for cluster members, but in all likelihood more appropriate for the foreground population being probed.}. By imposing high photometric quality, we are more likely to select objects that are in fact of early type (although, as discussed earlier, there can be some degree of contamination by dusty AGBs, especially among the reddest objects), but we are restricting our sample to intrinsically bright stars as we move to higher distances. We obtain a list of 1972 sources, reaching $K_{{\mathrm S}}=13.6$, which is the magnitude of a late-B main sequence star in Wd~1. This list is then cross-matched with \textit{Gaia}~EDR3 and the \citet{bj21} list of inferred distances\footnote{Although the simple prior used by \citet{bj21} does not permit us to trust distances to individual stars, here we use them to infer bulk properties of the population, a task for which they provide a satisfactory approximation.}. In Fig.~\ref{extinction_run}, we plot the observed $(BP-RP)$ against the distances from \citet{bj21}. Only objects with $BP<21$ are included, as fainter sources lack reliable colours, resulting in a total of 1721 sources.

Since most stars selected will be of early type, the observed colour should be an approximate measurement of the extinction (bar the intrinsic colour of the star, which ranges from $-0.2$ to $-0.5$~mag from late-B to early-O stars). For distances lower than $\sim1\:$kpc, all stars have low extinction. Around $d\sim1\:$kpc, the extinction suddenly grows, with no values lower than $(BP-RP)\approx\ 0.5$ found any more, and values above $(BP-RP)\approx\ 1.5$ appearing for the first time. There is an abrupt change in the distribution at $d\approx2.8\:$kpc. The number of sources in the graph abruptly decreases, while stars with $(BP-RP)<2$ almost disappear. Given the selection procedure, in particular the high photometric quality required, the only sensible interpretation is the presence of an extinction wall at this distance. This is corroborated by the abrupt increase in the average colour between 2.6 and 3.0~kpc as seen in the running average displayed in Fig.~\ref{extinction_run}). The behaviour of the ($J-K_{\mathrm{S}}$) colour is identical, although the range of values is much smaller, as expected. Comparison to recent models of galactic structure \citep[see, e.g.][]{Reid2019, hou21} suggests that this extinction wall at $\sim2$.8~kpc is associated with molecular clouds located in the Scutum-Centaurus arm, while the smaller rise in extinction around 1~kpc is due to the Sagittarius arm, which is not very prominent along this sightline. Interestingly, the distance to the cluster obtained by \citet{aghakan21} coincides with the distance to the extinction wall. 
 
 \begin{figure}
   \resizebox{\columnwidth}{!}
   {\includegraphics[angle=-90, clip]{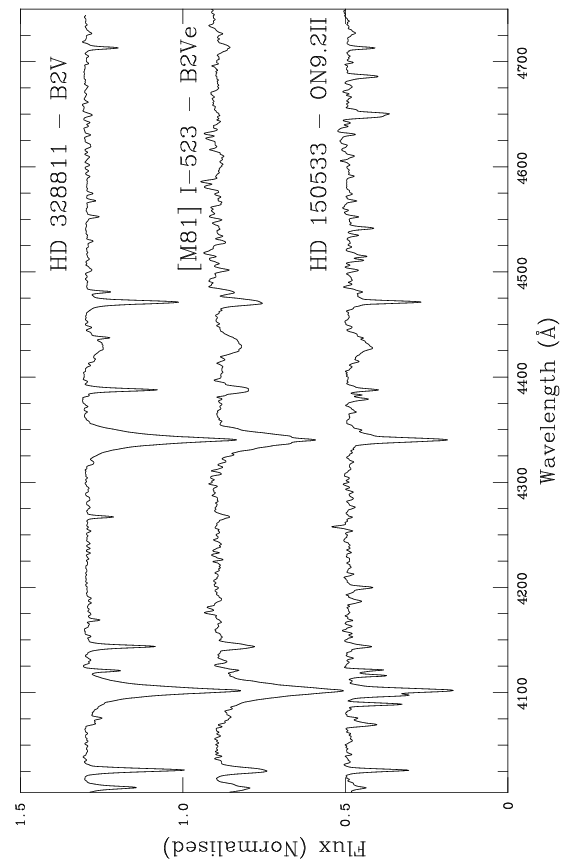}}
   \caption{Spectra of stars in the foreground to Wd~1 whose extinction is sufficiently low to allow  blue spectroscopy. }
   \label{fores}
\end{figure} 
 
The distribution of extinction may be further probed with our observations and the characteristics of known clusters in the region. Figure~\ref{fores} shows the classification spectra of a few objects that could be observed with the blue arm of AAOmega. We classify HDE~\num{328811}, whose EDR3 distance is $1.6$~kpc \citep{bj21}, as B2~V. It has $UBV$ photometry from \citet{whiteoak63}, who estimates $E(B-V)=0.5$. HD~\num{150533} is listed as a Be star in SIMBAD, despite being classified as an O9~Ib supergiant by \citet{garrison77}. We find a slightly lower luminosity for this object, whose EDR3 distance is $2.5\:$kpc \citep{bj21}. $UBV$ photometry is available from a number of sources, starting with \citet{whiteoak63}, and indicates $E(B-V)=1.0$. \citet{cantat20} identify a number of moderately distant clusters around our field: NGC~6216 ($d=2.6\:$kpc, $A_{V}=2.15$), Ruprecht~121 ($d=2.0\:$kpc, $A_{V}=2.54$), UBC~548 ($d=2.4\:$kpc, $A_{V}=2.27$) and UBC~669 ($d=2.4\:$kpc, $A_{V}=1.9$). All these observations confirm the presence of patchy and relatively low extinction out to around $2.5\:$kpc.

Contrarily, the distances to the high-reddening blue sources identified with our AAOmega observations (see Table~\ref{tab:2dFblue}) are above the discontinuity observed. Unfortunately, all these sources are very faint, and their parallax measurements are subject to large uncertainties. However, only one object, \#365, has a nominal distance below $2.8\:$kpc, namely $2.0^{+0.7}_{-0.3}\:$kpc \citep{bj21}, although, with $\pi/e_{\pi}=4.6$, this value is highly uncertain. Interestingly, many sources with $4\la(BP-RP)\la5$ have nominal distances in the vicinity of $3\:$kpc. Of particular interest is a close group of three early-B supergiants, \#236, \#255 and \#263, within less than $1\arcmin$ of each other and having essentially the same proper motions (within their errors), which are probably signposting a young open cluster or association, about $25\arcmin$ north of Wd~1.

\begin{figure}
   \resizebox{\columnwidth}{!}
   {\includegraphics[angle=0, clip]{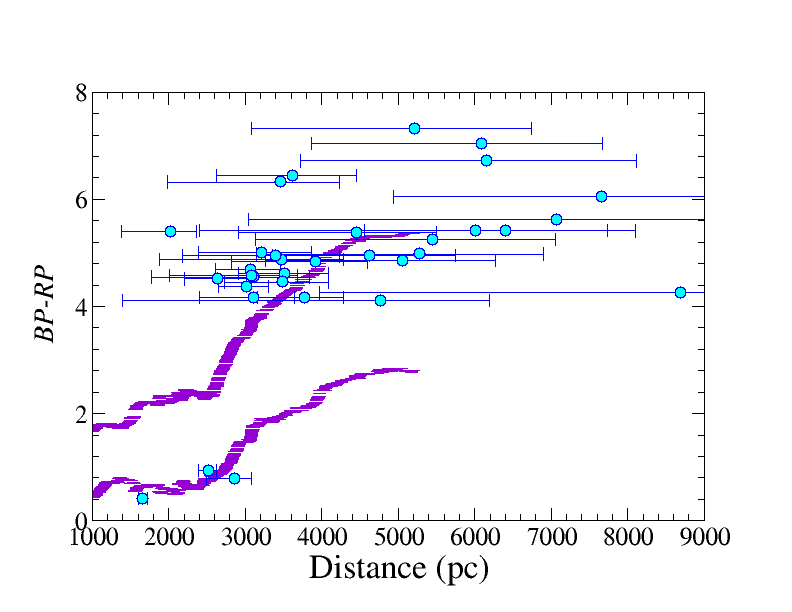}}
   \caption{As in Fig.~\ref{extinction_run}, but for the sample of "blue" stars observed spectroscopically (cluster members are not included). The thick discontinuous lines represent the running standard deviation from Fig.~\ref{extinction_run}, which suggests that the distances in \citet{bj21} for our sample may be underestimated. }
   \label{extinction_spectra}
\end{figure} 

\subsection{Cluster radial velocity}
\label{subsec:radvels}

The brightest members of Wd~1 are extremely luminous supergiants with complex atmospheric structures. Many of them show substantial spectroscopic variability, ranging from line-profile changes -- indicative of pulsation -- to changes in spectral types by several subtypes \citep{ritchie09bins,clark10}. This results in rather significant variations in the RV measured at different epochs \citep{ritchie09bins}. Furthermore, some of them can even show substantial differences in the RV measured from lines corresponding to different ions, as is the case of the yellow hypergiant Wd1-W265 \citep{ritchie09bins}. Because of this, defining an average cluster RV value is very difficult, unless a large sample is observed at a given time or -- even preferably -- over a large number of epochs.

\textit{Gaia}~DR2 provides radial velocities for five bright cluster members, the YHGs Wd1-W4, Wd1-W8a, and Wd1-W265, and the RSGs Wd1-W26, and Wd1-W75. Except for Wd1-W4, all the RVs have large dispersions, reflecting the variability. They are, however, long-term averages and therefore quite reliable. The RSG Wd1-W75 is not part of our \textit{Gaia}-based sample, as its RUWE $>1.4$ removed it from the selection process. The weighted (full weights) average heliocentric radial velocity for the four objects in the \textit{bona fide} sample is $-47.4 \pm 0.4\:\mathrm{km}\,\mathrm{s}^{-1}$.  After the solar motion correction \citep{EACC2006}, the average radial velocity with respect to the Local Standard of Rest is $\nu_{\mathrm{LSR}}= -43.0 \pm 0.4 \:\mathrm{km}\,\mathrm{s}^{-1}$. If weight is not taken into account, the heliocentric mean value is $-45.4 \pm1.6\:\mathrm{km}\,\mathrm{s}^{-1}$. Including Wd1-W75 in the calculation of the cluster radial velocity changes the unweighted mean by  
about $-1\:\mathrm{km}\,\mathrm{s}^{-1}$, bringing it closer to the weighted mean of $-47.4\:\mathrm{km}\,\mathrm{s}^{-1}$. Although the \textit{Gaia} RV values are almost certainly very good estimates of the stellar average radial velocities, the formal errors likely underestimate our uncertainty in the definition of a radial velocity for the cluster. Based on the dispersion of individual RVs, we take $\pm4.0\:\mathrm{km}\,\mathrm{s}^{-1}$ as an upper limit on this uncertainty.

From our AAOmega data, we obtain single-epoch RVs for six cluster members, the four RSGs plus Wd1-W4 and Wd1-W265. Their average is $-48.7\:\mathrm{km}\,\mathrm{s}^{-1}$, in excellent agreement with the \textit{Gaia}~DR2 value. Both values are also in good agreement with the RVs measured by \citet{ritchie22} from a sample of putatively non-binary early-type stars (which average $-43\:\mathrm{km}\,\mathrm{s}^{-1}$), although we cannot make a direct comparison to make sure that their velocities are in the same system as those from \textit{Gaia}. Given the correction to the Local Standard of Rest in this direction ($+4.1\:\mathrm{km}\,\mathrm{s}^{-1}$), we can conclude that the velocity of the cluster is not very far away from $v_{\mathrm{LSR}}=-43\:\mathrm{km}\,\mathrm{s}^{-1}$. This value is substantially higher than the $-55\:\mathrm{km}\,\mathrm{s}^{-1}$ \textit{assumed} by \citet{kothes07}, and, according to the Galactic rotation curve in this direction \citep[e.g.][]{Reid2019}, would place the cluster at a very low distance ($\la3\:$kpc).

However, the gas emission maps displayed by \citet[e.g. their figure 3]{Reid2019} very strongly suggest that clouds in this direction do not follow closely the rotation curve. On the other hand, a combination of recent studies of various spiral tracers, including ethanol and water masers, draws a $l-\nu_{\mathrm{LSR}}$ map \citep{hou21} where the radial velocity obtained for Wd~1 would place it very close to the Norma-Outer arm\footnote{For easy visualisation of the distribution of radial velocities in this direction and the difficulty in identifying tracers of different arms, see fig.~1 in \citet{colombo22}}. Moreover, the distance that we derive for the cluster also agrees with estimates to the arm in this direction. Given that the extinction wall around 2.8\,--\,3.0~kpc is in all likelihood associated with the Scutum-Crux arm, Wd~1 most likely lies in the Norma arm, probably on its near side.

\subsection{The field around Westerlund~1: a hive of RSGs }
    
The RSGs in Wd~1 are extremely bright near-IR sources, with Wd1-W26 and Wd1-W237 presenting $K\la2.5$, Wd1-W20 at $K\approx3.1$, and Wd1-W75 about 0.3~mag fainter \citep{borgman70}. Recently, \citet{beasor21} have argued that these objects represent an older burst of star formation than the bulk of the early-type stars. Although the assumption of a pretty homogenous extinction towards the cluster may play a major role in this conclusion, the possibility that the cluster is part of a larger association formed along a moderately long timespan is thus worth exploring. To this aim, we looked for stars of similar infrared brightness that could be RSGs related to the cluster in the field surrounding it.

There are 24 stars brighter than $K_{{\mathrm S}}=4$ in the (2 degree diameter) field that we observed. Of them, only two are brighter than Wd1-W26 and Wd1-W237, the catalogued Mira variable V823~Ara and the OH/IR star HDE~\num{328913}, which we observed (our target \#3), finding that it is a very luminous star, which we classify as M4.5\,I. This is the only one among these 24 objects whose proper motions suggest it could have been ejected from the cluster (although it is almost one degree away and a more careful analysis would be necessary to ascertain this possibility). However, its EDR3 parallax, $\varpi=0.58\pm0.04$~mas seems to completely rule out this possibility. In fact, only 5 of the 24 sources with $K_{{\mathrm S}}<4$ have a \citet{bj21} distance $\ga3\:\mathrm{kpc}$. These four objects have proper motions roughly in agreement with the cluster averages, but they are all located $>20\arcmin$ away from the cluster centre. We have a spectrum for one of them,  \#390, an M3.5\,Ib supergiant with $RV=-50\:\mathrm{km}\,{\mathrm s}^{-1}$. Another object of interest for which we have a spectrum is \#152, with $K_{\mathrm{S}}=3.2\pm0.3$ . This star has essentially the same ppms as the cluster, but its parallax $\varpi=0.54\pm0.14$ places it more than $2\,\sigma$ away from the cluster average. However, given that $\varpi/e_{\varpi}$ is only $4.0$, its high luminosity (M3\,Ia) 
and a radial velocity similar to those of cluster members ($RV=-44\:\mathrm{km}\,{\mathrm s}^{-1}$), we cannot fully rule out a connection with the cluster until a more accurate value of the parallax is available. In any event, this object is $>20\arcmin$ away from the cluster. 
    
\begin{figure}
   \resizebox{\columnwidth}{!}
   {\includegraphics[angle=0, clip]{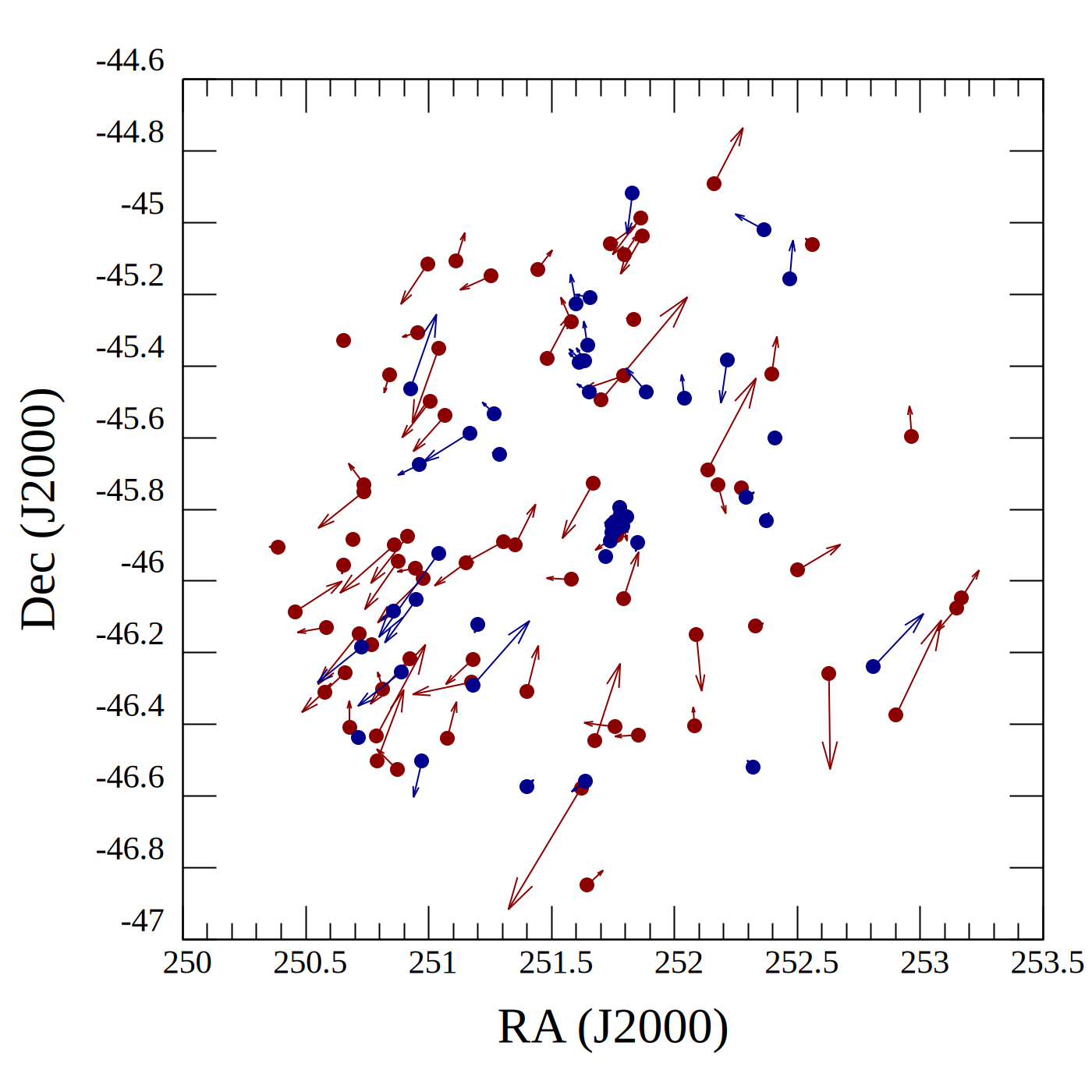}}
   \caption{Relative proper motions of massive stars observed with AAOmega. The arrows are proportional to the difference between the proper motions of the star and the average proper motions of the cluster. Blue dots represent our "blue" targets. Red dots are objects that we classify as red supergiants. }
   \label{relative}
\end{figure}

As we move towards fainter near-IR magnitudes, we fail to find the combination of high luminosity, RV and proper motion that would allow us to claim a connection to the cluster. In Fig.~\ref{relative}, we carry out a simple experiment, by subtracting from the proper motions of all the RSGs that we have observed the cluster average values found above. As can be seen, many of the RSGs have proper motions differing significantly from those of the cluster (although we must note that errors are in some cases very large). The very few objects for which this difference in proper motions might be compatible with an ejection from the cluster are very faint in $K_{\mathrm{S}}$ (e.g.\ the K1\,Ia star \#214, with $K_{\mathrm{S}}=6.82\pm0.02$) or not very luminous (e.g. the M1\,Ib object \#248 or the M2\,Ib \#334). Therefore, we find little to no evidence of RSGs ejected from the cluster either in our spectroscopic sample or in the sample of stars brighter than $K_{\mathrm{S}}=4$. Likewise, we find no evidence for an extended population surrounding the cluster that could be indicative of a large OB association centred on the cluster \textit{\`a la} Per~OB1 \citep[see, e.g.][]{deburgos20}. A few of the RSGs and some of the blue stars have proper motions and parallaxes that agree very well with those of Westerlund~1, and may be indicating the presence of a diffuse OB association with a range of ages, although they could also be part of the general population of massive stars associated with the Norma arm. In any event, their presence very strongly suggests that the proper motions of Westerlund~1 are indistinguishable from those of stars in its neighbourhood. Since they are also quite similar to those of the much more numerous population associated with the foreground Scutum-Crux arm, we are tempted to conclude that proper motions in this direction mainly reflect the relative motion between stars following their Galactic orbits and the Sun.
    
Despite this lack of connection to the cluster, the large population of RSGs found is remarkable in itself. Among the hundreds of red luminous stars observed, close to 40 have spectral classifications compatible with being evolved massive stars. As a comparison, a circle of radius $1\degr$ centred on $\chi$~Per (i.e. the densest region of Per~OB1) contains nine RSGs. Such a bountiful harvest of RSGs is testimony to the richness of the Scutum-Crux arm in this direction. Within the small field covered by our AAOmega observations, we find a very substantial population of massive stars, evidence for ongoing star formation and thick dark clouds that obscure our view to Westerlund~1.
    
\section{Conclusions}

Our analysis of \textit{Gaia} EDR3 data in a circle of radius $12\arcmin$ surrounding Westerlund~1 reveals the extraordinary complexity of this field. We find extreme crowding, huge variations in foreground extinction and a very complex kinematic configuration. As a consequence, the direct application of a non-parametric method to the positional and kinematic data of stars does not provide us with a clean sample of cluster members, but needs further pruning based on the distance to the cluster centre and the $(BP-RP)$ colour of the stars. Selecting the reddest stars ($(BP-RP)\ >$\ 4) and restricting ourselves to the central $3\farcm5$ of the circle, we find that:

\begin{enumerate}
      \item The population associated with the cluster is well separated. The Bayesian membership analysis leads to 401 bona fide cluster members, 142 of them with parallaxes corrected from EDR3 systematic zero-points. This sample includes the majority of spectroscopically catalogued members in \citet[]{clark20}.
      \item The parallax of the cluster is estimated to be $\varpi_{c, t} = 0.238 \pm 0.012$~mas. This value is statistically indistinguishable from that obtained by \citet{beasor21} for a selection, based on proper motion analysis, of catalogued OB members with data in \textit{Gaia}~EDR3 ($\varpi_{c, t} = 0.243 \pm 0.027$~mas). Different selection methods and samples of Wd~1 members lead to the same results for the cluster parallax, using the best available astrometric data. Nevertheless, we note that the parallax corrections for cluster members are generally quite extreme (typically, $-0.08$~mas), and therefore systematic effects cannot be completely ruled out. The analysis presented by \citet{maiz22} suggests that a large systematic is unlikely for the range of colours and magnitudes found in the cluster, but this needs to be confirmed. A distance of 4.2~kpc likely places Wd~1 in the Norma arm.
      \item The cluster distributes on the plane of the sky (RA, DEC) with an elliptical shape, where the principal diameter is $\approx7$~arcmin. This value must be considered an upper limit  on the cluster size, which would lead to a major semiaxis of 3.5~arcmin, well above the radius determined by other authors \citep[see, for example,][who give a value $R= 1\farcm2$]{Dias2002}. The radius of a cluster is a fuzzy concept that admits different definitions and approaches \citep[i.e.][]{sanchez20}. If we assume that the minor axis of the ellipse is $2\farcm5$, the effective radius would be close to $3\arcmin$, which at the distance estimated for Wd~1 corresponds to 3.7~pc.
       \item The proper motion of the cluster derived from the 401 \textit{bona fide} cluster members is $\mu_\alpha(c) =$\ $-2.231$ and $\mu_\delta(c) =$\ $-3.697$~mas/a, which at the distance of 4.2 kpc, yields $V_\alpha(c)$  and $V_\delta(c)$ space velocities of $44.5\pm 3.2$  and $73.8\pm 5.3\:\mathrm{km}\,\mathrm{s}^{-1}$, respectively.
      \item There are indications of the presence of a low-density halo surrounding the cluster, as many of the very red stars plotted in Fig.~\ref{fig:XY_RB} have proper motions compatible with those of the cluster (see Fig.~\ref{fig:VPD_RB}). However, we also find stars with fully compatible astrometric parameters at much higher angular distances (for instance, stars \#246, \#359, \#389, \#395, and \#421 in Table~\ref{tab:2dFblue} have parallax and proper motions fully compatible with membership within their errors, but are all located $>20\arcmin$ away from the cluster). This suggests that the proper motions of Wd~1 are not distinguishable from those of the field population in the Norma arm with the current level of accuracy. More accurate proper motions, such as those in DR4 (perhaps helped by spectroscopy), will be needed to prove the existence of a halo and search for runaway stars ejected from the cluster.    
      \item \textit{Gaia}~DR2 provides radial velocity values for four stars considered as members of the cluster. The weighted average of the radial velocities provides different mean values depending on the chosen weights, in such a way that we adopt a heliocentric radial velocity of $-45.4\:\mathrm{km}\,\mathrm{s}^{-1}$ with an internal precision of $1.4\:\mathrm{km}\,\mathrm{s}^{-1}$, corresponding to the unweighted average estimation. This may be an underestimation of the true uncertainty in the RV value,  despite the quality of \textit{Gaia} data, given the small number of objects for averaging, and the large number of RV variable stars. This value is fully consistent with the average RV of 58 OB stars without convincing signs of binarity \citep[$-43.1\:\mathrm{km}\, \mathrm{s}^{-1}$;][]{ritchie22}.
     
      \item Many of the astrometric members detected have colours indicative of higher reddening than most of the members listed by \citet{clark20}. The majority of these objects are located very close together in the sky, in a region to the east of the cluster centre and a strip that extends between the two main groupings seen in optical images. About 15 of these very red objects have infrared magnitudes and colours that identify them as luminous OB supergiants. This finding implies that extinction across the face of Wd~1 is much more variable than previously assumed. As a consequence, the average extinction and its dispersion are likely higher than found by \citet{damineli16}, in agreement with \citet{andersen17}.
      
      \item The sightline to Wd~1 goes through an extremely rich portion of the Scutum-Crux arm, which contains a huge population of massive stars (see Appendix~\ref{app:omega}). The distribution of sources in this direction suggests that this population is mostly located at a distance $\sim3\:$kpc, behind a wall of dark clouds that contribute substantially to the extinction to the cluster.  
   \end{enumerate}
In all, \textit{Gaia} EDR3 data provide us with a much improved view of the cluster properties. Future data releases will likely sharpen this view.

\begin{acknowledgements}
This paper is dedicated to the memory of Simon Clark, who tragically passed away as this work was beginning. For twenty years he and the first author, together with many others, explored Westerlund~1. The expedition is far from over, but the captain is no longer at the helm. All new crews that take up this task will be sailing on his trail.\\
We thank the anonymous referee for helpful suggestions.
    This research is partially supported by the Spanish Government Ministerio de Ciencia, Innovaci\'on y Universidades under grants PGC2018-\num{093741}-B-C21/C22 and PGC2018-\num{095049}-B-C21/C22 (MICIU/AEI/FEDER, UE) and the Generalitat Valenciana under grant PROMETEO/2019/041. EJA acknowledges financial support from the State Agency for Research of the Spanish MCIU through the "Center of Excellence Severo Ochoa" award to the Instituto de Astrofísica de Andalucía (SEV-2017-0709)”. RD acknowledges further support from the Spanish Government Ministerio de Ciencia e Innovaci\'on through grant SEV 2015-0548, and from the Canarian Agency for Research, Innovation and Information Society (ACIISI), of the Canary Islands Government, and the European Regional Development Fund (ERDF), under grant with reference ProID\num{2017010115}. The AAOmega observations have been supported by the OPTICON
project (observing proposal 2011A/014), which was
funded by the European Commission under the Seventh Framework Programme (FP7). \\
    This research has made use of the Simbad, Vizier and Aladin services developed at the Centre de Donn\'ees Astronomiques de Strasbourg, France. This work has made use of data from the European Space Agency (ESA) mission
{\it\textit{Gaia}} (\url{https://www.cosmos.esa.int/gaia}), processed by the {\it\textit{Gaia}}
Data Processing and Analysis Consortium (DPAC,
\url{https://www.cosmos.esa.int/web/gaia/dpac/consortium}). Funding for the DPAC
has been provided by national institutions, in particular the institutions
participating in the {\it\textit{Gaia}} Multilateral Agreement.
The {\it Gaia} data are processed with the computer resources at Mare Nostrum and the technical support provided by BSC-CNS.
In this work, we have made extensive use of TOPCAT \citep{TOPCAT2005}. We thank its author, and subsequent contributors, for the creation and development of this tool.  

\end{acknowledgements}

\bibliography{clusters,gaia,rsgs,bins,own}

\begin{appendix}

\section{Is there a cluster in front of Westerlund~1?}
\label{app:bh197}

The nature of the BG is an open question. 
The open cluster vandenBergh-Hagen~197 \citep[MWSC~2458; BH~197;][]{vdbh75} is catalogued to lie in the Northern part (in Galactic coordinates) of the area occupied by the BG. Catalogue data for this object show disparate values for the basic physical properties. While the two studies that have recently estimated the distance and age of this cluster coincide \citep{kharchenko13,dias14} in placing it between 1.6 and 2~kpc and assign a similar extinction, $E(B-V) \sim1$, they differ widely in the estimation of age: \citet{kharchenko13} dates it close to 1~Ga, 
while \citet{Dias2002} give 30~Ma, which seems more compatible with the colour observed for the BG. With this information, we can speculate that the area west of Wd~1, at least within the radius of $12\arcmin$ chosen for this study, may contain a star formation region with a minimum size of 9~pc, which includes the cluster BH~197 surrounded by a halo of low-density blue stars at an average distance of 2~kpc.

The membership analysis performed by \citet{sampedro17}, based on UCAC4 \citep{zacharias13} data, and on the previous work by \citet{dias14}, selects a set of 274 members (considered as such by the four methods used in that work) within a radius of $4\arcmin$ around the coordinates of the cluster centre given by \citet{kharchenko13}.

Of the 274 BH~197 members catalogued by \cite{sampedro17}, 253 are in our initial catalogue. The $G$\ vs\  $(BP-RP)$ CMD and proper-motion VPD drawn with $Gaia$~EDR3 data for this sample show distributions far away from those expected for a stellar cluster. The CMD presents different sub-groupings at different colour ranges that cannot be explained by a stellar system with a single stellar population, even advocating a very complicated extinction pattern. The same can be said for its distribution in proper motion space. Members span more than 10 mas/a along the main axis, and are far from resembling what is expected for a star cluster at the distance of 2~kpc.  Thus, we conclude that BH~197 cannot be considered an open cluster in the classical sense of the term, but rather, in the best event,  a subregion of a larger star-forming region that could extend beyond 9 pc, this  lower limit being defined by the sampling radius of our data.

The distribution of YSOs along the line of sight, which crosses the Galactic plane in the direction $l\ \approx$\ 340$\degr$, shows a clumpy pattern \citep{kuhn2021}, with groupings of YSOs (identified from \textit{Spitzer}'s catalogue) associated with the Carina-Saggitarius and Scutum-Centaurus arms, but with some clustering in the inter-arm region. Our BG appears to be located in the vicinity of these concentrations and could represent the more evolved optical counterpart of the ongoing star-formation process. There is an additional argument that supports the hypothesis that the BG is a relatively extensive, active star-forming region.  Dense-gas hub-filament systems (HFSs) appear to be good tracers of high-mass star-formation activity \citep{schneider2012, peretto2013, kumar2020}. One HFS  (HFS14082) detected by \cite{kumar2020} is located within the area covered by the BG. This hub is at a distance compatible with that estimated for our Blue Group, especially if we take into account the different distance estimation methods and their associated uncertainties. 

\newpage

\section{Cluster members}
\label{app:members}

In Table~\ref{clarkmembers}, we list the members of Wd~1 with spectroscopic observations in the catalogue of \citet{clark20} and their \textit{Gaia} EDR3 basic information. The coordinates in \citet{clark20} were collected from a variety of sources, and had different accuracies. A majority (114) have accurate coordinates, and the nearest EDR3 source is within $0\farcs5$. A few (8) are $>1\arcsec$ away from the nearest EDR3, but all are within $1\farcs5$. The only exception is the RSG Wd1-W20 for which \citet{clark20} erroneously list the same coordinates as for Wd1-W237.

In addition to the members already known, we list the newly confirmed member W1070. This object is outside the $3\farcm5$ region where we have selected \textit{bona fide} members, but its astrometric values are typical of cluster members. Although, its spectrum is very poor, the narrow Paschen lines and presence of a very prominent \ion{C}{iii}~8502\,\AA\ line identify it as an O9\,I star.

\newpage

\setlength{\tabcolsep}{0.3em}
\begin{landscape}
\begin{table}
\caption{\textit{Gaia}~EDR3 data for previously catalogued members of Westerlund~1 (and one new member). Names are as in \citet{clark20}. Parallaxes are not corrected for zero point bias. The KM column indicates whether the star is considered a \textit{bona fide} kinematic member by our Bayesian analysis ($+$) or not ($-$). The AA$\Omega$ column indicates if that star was observed by this instrument. \textit{This table will be available in electronic format in the published version.}\label{clarkmembers}}
\centering
\begin{tabular}[h]{l c c c | c c c | c c c c | c c}
\hline\hline
\noalign{\smallskip}
ID & RA & DEC & Gaia EDR3 & Plx & pmRA  & pmDE & $G$ & $BP$ & $RP$ & $BP-RP$ &AA$\Omega$&KM\\
& J2000 & J2000 &  & (mas) & (mas$\:\mathrm{a}^{-1}$) & (mas$\:\mathrm{a}^{-1}$) & &&& & & \\
\noalign{\smallskip}
\hline
\noalign{\smallskip}
W1&16:46:59.4&$-45$:50:46.8&5940106758709791232& $0.12\pm0.06$ & $-1.86\pm0.07$ & $-3.77\pm0.07$ & $15.572\pm0.003$ & $18.482\pm0.014$ & $14.091\pm0.005$ &4.391& $-$ & $+$\\
W2a&16:46:59.7&$-45$:50:51.2&5940106758703247360& $0.26\pm0.04$ & $-2.02\pm0.05$ & $-3.41\pm0.05$ & $13.682\pm0.003$ & $16.659\pm0.010$ & $12.187\pm0.006$ &4.472& $-$ & $+$\\
W4&16:47:01.4&$-45$:50:37.3&5940106763014985088& $0.18\pm0.06$ & $-2.24\pm0.07$ & $-3.42\pm0.06$ & $10.938\pm0.003$ & $14.454\pm0.011$ & $9.496\pm0.007$ &4.958& $+$ & $+$\\
...&...&...&...&...& .... &... & .... & ... & ... &...& ... & ...\\
\noalign{\smallskip}
\hline
\end{tabular}

\end{table}
\end{landscape}		

\newpage
\section{Stars observed with AAOmega}
\label{app:omega}

In Tables~\ref{tab:2dFblue} and~\ref{tab:2dFred}, we list all the stars that were observed with AAOmega, divided according to their nature. Table~\ref{tab:2dFblue} contains blue stars and a handful of yellow supergiants. Many of these objects are quite faint, and their spectra are quite noisy (sky subtraction is worse in spectra with low signal-to-noise). For some, the only features seen are broad and shallow Paschen lines, which identify them as OB stars \citep[mid and late B stars have deeper Paschen lines; see][]{negueruela10Wd1}. In a few, a moderately strong \ion{C}{iii}~8502\,\AA\ line is seen, signalling a luminous star with spectral type close to O9.

Our targets include 9 B-type supergiants, 3 A-type luminous supergiants and five F-type supergiants. This is a surprisingly high number, considering the small field surveyed. The high number of supergiants observed indicates that we are only observing the intrinsically most luminous early-type stars. Most of them have a 2MASS ($J-K_{\mathrm{S}})$ colour $>1.5$, indicative of very high extinction -- the intrinsic colour of all these objects is close to $(J-K_{\mathrm{S}})\approx0$, with only the F-type supergiants approaching 0.3. A few of the most luminous supergiants have colours $(J-K_{\mathrm{S}})>2.0$, suggesting, and parallaxes compatible with being, background objects behind the Norma arm.

Table~\ref{tab:2dFred} contains cool stars (with spectral types G and later, corresponding to our "red" targets). Targets were ordered by their brightness in the $i$ or $I$ band. This is reflected in the quite strong correlation between ID number and $G$ band brightness. Spectral types have been estimated following the criteria discussed in \citet{negueruela12}. The quality of the spectra differs greatly, depending on many factors. For average count rates below about 1000, sky subtraction becomes substantially worse, and many of the fainter targets only have approximate classifications, or simply an M classification that indicates that TiO bands are visible, but the quality of the spectrum prevents further classification.

\subsection{Some interesting early-type stars}

Star \#412 is coincident with the mid-IR source IRAS~16475$-$4609. Based on the morphology of the WISE images and presence of the nearby candidate young stellar object IRAS~16474$-$4610, \# 412 could be the ionising star of a small \ion{H}{ii} region at a very high distance.

Star \#168 is a luminous B5\,Ia supergiant (see Fig.~\ref{brightomega}). It is located approximately $2\farcm5$ to the north of the compact group formed by \#236, \#255 and \#263, and its proper motions, though similar to those of the other three, are compatible with having travelled from their vicinity. All these stars have parallaxes compatible with a distance $\sim3\:$kpc.

\begin{figure}
   \resizebox{\columnwidth}{!}
   {\includegraphics[clip]{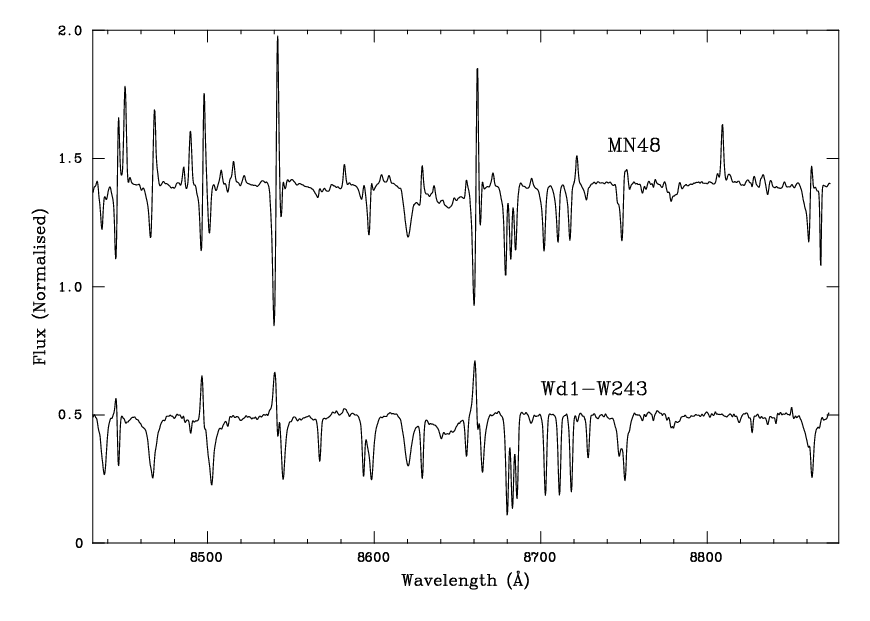}}
   \caption{AAOmega spectrum of the LBV MN48, compared to that of Wd1-W243. In both stars, the strongest absorption lines correspond to \ion{N}{i}. The Paschen lines in Wd1-W243 are almost entirely in absorption, while those in MN48 display P-Cygni profiles. See text for a description.
   \label{lbvs}}
\end{figure}

Target \# 119 is the central star of the ring nebula [GKF2010] MN48 \citep{gvaramadze10}. These authors proposed it as a candidate LBV star. \citet{wachter10} classified it as an unusual emission-line star, based on $H$- and $K$-band spectra. \citet{kniazev16} presented comprehensive photometry and spectroscopy of the source, confirming its LBV nature. Based on the evolution of brightness and colours, \citet{kniazev16} argued that the star become cooler and more luminous between 2009 and 2011, and then faded as its temperature increased again. If this interpretation is correct, our spectrum was taken close to maximum light. The spectrum is shown in Fig.~\ref{lbvs}, compared to that of the known cluster LBV Wd1-W243. The $I$-band spectrum of Wd1-W243 is very similar to those reported for 2005\,--\,2009 by \citet[their fig.~11]{ritchie09W243}, although the emission cores of the Paschen lines seem to have weakened, an effect that is more clearly seen in Pa~11, which is essentially a pure absorption profile now. The absorption spectrum is dominated by \ion{N}{i} lines, which correspond well with a 8\,500~K hypergiant model \citep[their fig.~14]{ritchie09W243}. 

The spectrum of MN48 is dominated by much stronger emission lines. The Paschen lines show weak P-Cygni profiles. The \ion{Ca}{ii} lines have much stronger P-Cygni profiles, although at this resolution it cannot be discarded to attribute the absorption profiles to the broader Paschen lines. The \ion{N}{i} spectrum in absorption is somewhat weaker than in Wd1-W243, but still compatible with an early-A super/hypergiant. A low-resolution optical spectrum taken two months before our observations is dominated by emission lines corresponding to singly ionised metals, mainly \ion{Fe}{ii} \citep{kniazev16}. According to these authors, the star became noticeably hotter over the following four years. A spectrum taken in 2015 shows the upper Paschen lines essentially in absorption, while the \ion{O}{i}~8446\,\AA\ line, powered by Lyman $\beta$ fluorescence, remains strong (peaking around 1.4 times the continuum level, as in our spectrum). Our spectrum also shows a strong \ion{Fe}{ii}~8490\,\AA\ line, which is pumped by Lyman $\alpha$.

The proper motions of MN48 are almost identical to the cluster averages, but -- given the moderate errors -- may indicate a small difference around 0.3~mas in pmRA. As the star is located $\sim25\arcmin$ to the east and $\sim9\arcmin$ to the north, travelling this distance would require around 5~Ma, which makes association with Wd~1 possible. Given the range of ages currently being considered for Wd~1, the star should have been ejected early in the history of the cluster. \citet{kniazev16} measure an RV\,$=\,-37\:\mathrm{km}\,\mathrm{s}^{-1}$, which is also compatible with a slow ejection from Wd~1. It is worthwhile commenting that \citet{gvaramadze18} identifies another LBV (MN44) that may also have been ejected from Wd~1 (in approximately the opposite direction) and that also requires a long ($\sim4\:$Ma) travel time. If these associations are confirmed by more accurate proper motions, they would imply that both stars were ejected early in their lives, when they were O-type main sequence stars, and evolved into LBVs close to their current locations.

\begin{landscape}
\begin{table}
\caption{"Blue" targets observed with AAOmega. The IDs are used to identify targets in the text. \textit{This table will appear in electronic format only in the published version.}} \label{tab:2dFblue}
\centering
\begin{tabular}{lccccccccccc}
\hline
\noalign{\smallskip}
ID & \textit{Gaia} EDR3 ID & RA & DEC & Parallax & pmra & pmdec & $G$ & $d$ & Spectral &  $K_{{\mathrm S}}$ & $(J-K_{{\mathrm S}})$  \\
&&(ICRS)&(ICRS)&(mas)&(mas$\:\mathrm{a}^{-1}$)&(mas$\:\mathrm{a}^{-1}$)&&(kpc)&type&&\\
\hline
\noalign{\smallskip}
B01 & 5943070496608125952 & 16:44:43.6 & $-46$:17:22.2& $0.59\pm0.03$ & $0.06\pm0.03$ & $-1.92\pm0.02$ & $10.8$ & $1663_{-59}^{+60}$ & B2\,V &  10.2 & 0.1 \\
B02 & 5942702258975015296 & 16:43:53.3 & $-46$:30:07.8& $0.32\pm0.03$ & $-2.57\pm0.04$ & $-4.71\pm0.03$ & $12.1$ & $2857_{-220}^{+357}$ & B2\,Ve&  10.6 & 0.4 \\
B03 & 5940127894238886144 & 16:50:254 & $-46$:09:11 & $0.95\pm0.02$ & $0.60\pm0.02$ & $-3.96\pm0.02$ &10.4 & $1019_{-17}^{+16}$ & Bep &  6.7 & 2.1 \\
B04 & 5943178729790777216 & 16:43:42.3 & $-45$:27:45.5& $0.37\pm0.02$ & $-1.17\pm0.03$ & $-1.63\pm0.02$ & $9.2$ & $2516_{-109}^{+127}$ & ON9.2\,III &  7.8 & 0.2 \\
119 & 5940216130049700480 & 16:49:37.7 & $-45$:35:59.3& $0.16\pm0.07$ & $-1.92\pm0.09$ & $-3.69\pm0.07$ & $12.6$ & $5057_{-1210}^{+1799}$ & LBV &  5.4 & 1.8 \\
130 & 5943220678706033280 & 16:46:23.7 & $-45$:13:32.4& $0.29\pm0.06$ & $-2.45\pm0.08$ & $-2.88\pm0.06$ & $12.5$ & $3488_{-597}^{+764}$ & F8\,Ia &  6.4 & 1.5 \\
168 & 5943218995078596608 & 16:46:35.0 & $-45$:20:32.0& $0.28\pm0.05$ & $-2.38\pm0.06$ & $-3.03\pm0.05$ & $13.1$ & $3017_{-285}^{+368}$ & B5\,Ia &  7.3 & 1.4 \\
185 & 5943125296078454272 & 16:45:09.6 & $-45$:38:46.1& $0.27\pm0.05$ & $-2.55\pm0.07$ & $-3.65\pm0.06$ & $13.5$ & $3071_{-388}^{+464}$ & B0\,I &  7.3 & 1.5 \\
221 & 5940057937824693504 & 16:46:33.2 & $-46$:33:32.1& $0.24\pm0.09$ & $-2.80\pm0.11$ & $-3.99\pm0.08$ & $13.9$ & $3457_{-770}^{+1473}$ & B5\,Ia &  6.1 & 2.0 \\
234 & 5940060239927292544 & 16:45:36.2 & $-46$:34:20.5& $0.14\pm0.06$ & $-1.97\pm0.08$ & $-3.52\pm0.06$ & $14.2$ & $4631_{-1120}^{+1278}$ & B2\,I &  7.9 & 1.5 \\
236 & 5943217453188230528 & 16:46:26.6 & $-45$:23:23.8& $0.25\pm0.06$ & $-2.63\pm0.08$ & $-3.43\pm0.06$ & $14.2$ & $3216_{-646}^{+822}$ & $\sim$B0\,I & 7.9 & 1.5 \\
246 & 5942701576100661376 & 16:42:51.4 & $-46$:26:07.2& $0.20\pm0.04$ & $-2.10\pm0.06$ & $-3.54\pm0.05$ & $14.4$ & $3781_{-506}^{+616}$ & F0\,I &  8.8 & 1.3 \\
255 & 5943217345811090048 & 16:46:28.9 & $-45$:23:04.9& $0.34\pm0.06$ & $-2.72\pm0.06$ & $-3.38\pm0.05$ & $14.4$ & $2638_{-380}^{+428}$ & $\sim$B0\,I & 8.7 & 1.3 \\
263 & 5943217384492082944 & 16:46:32.4 & $-45$:23:03.6& $0.22\pm0.05$ & $-2.57\pm0.06$ & $-3.35\pm0.05$ & $14.5$ & $3519_{-572}^{+608}$ & $\sim$B1\,I & 8.5 & 1.4 \\
266 & 5942757651197863296 & 16:42:54.1 & $-46$:11:05.1& $0.02\pm0.10$ & $-4.01\pm0.12$ & $-4.68\pm0.10$ & $14.5$ & $6404_{-1703}^{+1846}$ & A0\,Ia &  7.7 & 1.6 \\
273 & 5943213708000014208 & 16:46:36.6 & $-45$:28:21.5& $0.30\pm0.09$ & $-2.73\pm0.11$ & $-3.47\pm0.09$ & $14.7$ & $3079_{-601}^{+1072}$ & $\sim$B0\,I & 8.6 & 1.4 \\
302 & 5943170414734473344 & 16:43:50.7 & $-45$:40:31.6& $0.21\pm0.06$ & $-3.10\pm0.08$ & $-3.99\pm0.07$ & $14.9$ & $3920_{-680}^{+1102}$ & $\sim$F5\,I & 8.5 & 1.6 \\
313 & 5964250354726697984 & 16:49:52.6 & $-45$:09:22.9& $0.12\pm0.11$ & $-2.10\pm0.14$ & $-2.63\pm0.11$ & $15.0$ & $4451_{-1042}^{+1544}$ & F2\,Ia & 8.1 & 1.7 \\
323 & 5943087848276539008 & 16:43:47.8 & $-46$:03:08.2& $0.08\pm0.15$ & $-3.51\pm0.17$ & $-4.90\pm0.13$ & $15.0$ & $5209_{-1536}^{+2123}$ & F5\,Ia & 5.9 & 2.5 \\
353 & 5943088050119534208 & 16:43:25.7 & $-46$:05:00.3& $0.33\pm0.10$ & $-2.75\pm0.12$ & $-3.95\pm0.09$ & $15.8$ & $3108_{-737}^{+1332}$ & OB & 9.7 & 1.5 \\
359 & 5940188848416265344 & 16:49:29.8 & $-45$:49:51.9& $0.27\pm0.07$ & $-2.14\pm0.08$ & $-3.48\pm0.06$ & $15.7$ & $3109_{-540}^{+712}$ & O9\,III-V & 10.1 & 1.4 \\
360 & 5943243493572357888 & 16:46:37.7 & $-45$:12:31.7& $0.23\pm0.11$ & $-2.84\pm0.14$ & $-3.62\pm0.11$ & $15.7$ & $3396_{-839}^{+1213}$ & OB &  9.3 & 1.5 \\
375 & 5940231729371122048 & 16:48:09.6 & $-45$:29:22.5& $0.21\pm0.10$ & $-2.34\pm0.11$ & $-3.04\pm0.09$ & $15.5$ & $3618_{-835}^{+999}$ & $\sim$O9\,I & 7.3 & 2.2 \\
385 & 5964264029905952896 & 16:49:27.3 & $-45$:01:10.6& $0.48\pm0.10$ & $-3.40\pm0.14$ & $-3.26\pm0.11$ & $16.1$ & $2027_{-336}^{+651}$ & $\sim$B3\,Ib &  9.2 & 1.6 \\
389 & 5943077883952128768 & 16:44:48.0 & $-46$:07:14.5& $-0.11\pm0.10$ & $-2.36\pm0.12$ & $-3.94\pm0.10$ & $16.2$ & $7658_{-1936}^{+2724}$ & $\sim$08\,III--V & 9.0 & 1.8 \\
391 & 5943129389202734080 & 16:44:40.0 & $-45$:35:13.5& $0.00\pm0.13$ & $-4.11\pm0.18$ & $-4.51\pm0.14$ & $16.1$ & $5450_{-1600}^{+2311}$ & OB &  9.3 & 1.7 \\
395 & 5940190119718188672 & 16:49:09.9 & $-45$:46:00.9& $-0.03\pm0.14$ & $-1.90\pm0.16$ & $-3.55\pm0.14$ & $16.3$ & $6010_{-1721}^{+3603}$ & $\sim$O9 &  9.5 & 1.7 \\
398 & 5943095888455406976 & 16:44:09.7 & $-45$:55:20.7& $-0.12\pm0.17$ & $-4.66\pm0.21$ & $-6.04\pm0.17$ & $15.9$ & $6156_{-1962}^{+2440}$ & A3\,Ia & 6.4 & 2.7 \\
404 & 5940212243106889344 & 16:47:32.7 & $-45$:28:22.6& $-0.40\pm0.15$ & $-3.08\pm0.19$ & $-3.03\pm0.15$ & $16.7$ & $8685_{-2707}^{+4716}$ & OB &  10.6 & 1.3 \\
412 & 5940212243106889344 & 16:47:32.7 & $-45$:28:22.6& $-0.40\pm0.15$ & $-3.08\pm0.19$ & $-3.03\pm0.15$ & $16.7$ & $8685_{-2707}^{+4716}$ & OB &  10.6 & 1.3 \\
416 & 5940233236896477056 & 16:48:51.4 & $-45$:22:58.9& $-0.01\pm0.11$ & $-2.49\pm0.13$ & $-4.90\pm0.11$ & $16.2$ & $7069_{-2092}^{+4031}$ & OB &  9.1 & 1.7 \\
419 & 5943084721540120960 & 16:43:33.7 & $-46$:15:10.1& $-0.18\pm0.14$ & $-4.01\pm0.16$ & $-4.66\pm0.13$ & $16.0$ & $6090_{-1585}^{+2220}$ & A0\,Ia &  7.3 & 2.4 \\
421 & 5940026223790310784 & 16:49:16.8 & $-46$:31:07.3& $0.17\pm0.09$ & $-2.47\pm0.14$ & $-3.51\pm0.10$ & $16.4$ & $4770_{-1421}^{+3379}$ & $\sim$B3\,III--V &  10.7 & 1.3 \\
427 & 5943254561730974080 & 16:47:19.1 & -44:54:59.7& $0.12\pm0.11$ & $-2.45\pm0.14$ & $-4.86\pm0.11$ & $16.6$ & $5283_{-1609}^{+2136}$ & OB &  10.1 & 1.5 \\
\noalign{\smallskip}
\hline
\end{tabular}
\end{table}
\end{landscape}

\begin{landscape}
\begin{table}
\caption{"Red" targets observed with AAOmega. The IDs are used to identify targets in the text. Parallaxes are not corrected for zero point bias. The $v_{\textrm{hel}}$ column lists our measurements. Infrared magnitudes are from 2MASS and distance estimates, from \citet{bj21}. The radial velocities can be considered accurate to $\pm3\mathrm{km}\,\mathrm{s}^{-1}$ (the dispersion of their difference with respect to \textit{Gaia} DR2 values). \textit{Only the first three entries are shown. This table will appear in electronic format in the published version.}} \label{tab:2dFred}
\centering
\begin{tabular}{lccccccccccccc}
\hline
\noalign{\smallskip}
ID & \textit{Gaia} EDR3 ID & RA & DEC & Parallax & pmRA & pmDec & $G$ & $d$ & Spectral & $v_{\mathrm{hel}}$ & DR2 $v_{\mathrm{hel}}$ & $K_{{\mathrm S}}$ & $(J-K_{{\mathrm S}})$  \\
&&(ICRS)&(ICRS)&(mas)&(mas$\:\mathrm{a}^{-1}$)&(mas$\:\mathrm{a}^{-1}$)&&(kpc)&type&$\mathrm{km}\,\mathrm{s}^{-1}$&$\mathrm{km}\,\mathrm{s}^{-1}$&&\\
\hline
\noalign{\smallskip}
1 & 5940194586492682112 & 16:48:47.19 & $-45$:40:00.83 & 0.96$\pm0.04$ & $-$5.80$\pm0.04$ & $-$14.32$\pm0.04$ & 9.4 & 1004$_{-34}^{+30}$ & M4\:II & $-$20.8 & $-$18.8$\pm0.4$ & 4.5 & 1.4 \\
2 & 5943233877166753920 & 16:45:46.32 & $-45$:07:50.08 & 0.43$\pm0.03$ & $-$1.63$\pm0.03$ & $-$3.16$\pm0.02$ & 9.2 & 2144$_{-106}^{+129}$ & M0\:Iab & $-$43.3 & NA & 4.7 & 1.3 \\
3 & 5964268702827222656 & 16:48:38.64 & $-$44:53:29.37 & 0.58$\pm0.04$ & $-$1.05$\pm0.06$ & $-$2.14$\pm0.04$ & 6.9 & 1621$_{-111}^{+125}$ & M4.5\,I & $-$38.3 & $-$38.0$\pm0.5$ & 1.5 & 1.6 \\
&... &... & ...& ...& &... &... & ...& ...& ...& ...&... &... \\
\noalign{\smallskip}
\hline
\end{tabular}
\end{table}
\end{landscape}

\end{appendix}

\end{document}